\documentclass[9pt,twocolumn,journal]{IEEEtran}

\IEEEoverridecommandlockouts

\usepackage{color}
\usepackage{xcolor}

\usepackage{placeins}
\usepackage{amsthm}
\usepackage{amsmath}
\usepackage{mathtools}
\usepackage{amssymb}
\usepackage[named]{algo}
\usepackage[noend]{algpseudocode}
\usepackage{algorithm}

\usepackage{verbatim}
\usepackage{cite}
\usepackage{url}
\usepackage{cases}
\usepackage{upgreek}
\usepackage{bm}
\usepackage{nicefrac}

\ifCLASSOPTIONcompsoc
  \usepackage[caption=false,font=normalsize,labelfont=sf,textfont=sf]{subfig}
\else
  \usepackage[caption=false,font=footnotesize]{subfig}
\fi

\usepackage{cite}
\usepackage{balance}

\newtheorem{theorem}{Theorem}

\renewcommand{\IEEEQED}{\IEEEQEDopen}
\newdimen\origiwspc%
\newdimen\origiwstr%
\origiwspc=\fontdimen2\font% original inter word space
\origiwstr=\fontdimen3\font% original inter word stretch

\IEEEoverridecommandlockouts

\begin{document}

\title{A Cognitive Sub-Nyquist MIMO Radar Prototype}

%\author{\fontdimen2\font=0.5ex{Kumar Vijay~Mishra and Yonina C.~Eldar}\fontdimen2\font=\origiwspc% <-this % stops a space
\author{Kumar Vijay~Mishra, Yonina C.~Eldar, Eli Shoshan, Moshe Namer and Maxim Meltsin% <-this % stops a space
\thanks{K. V. M. is with The University of Iowa, Iowa City, Iowa, USA. E-mail: kumarvijay-mishra@uiowa.edu.}
\thanks{Y. C. E. is with the Weizmann Institute of Science, Rehovot, Israel. E-mail: yonina.eldar@weizmann.ac.il.}
\thanks{E. S., M. N., and M. M. are with the Andrew and Erna Viterbi Faculty of Electrical Engineering, Technion - Israel Institute of Technology, Haifa, Israel. E-mail: \{elis, namer, maxim.meltsin\}@ee.technion.ac.il.}% <-this % stops a space
\thanks{This project is funded from the European Union's Horizon 2020 research and innovation programme under grant agreement No. 646804-ERC-COG-BNYQ. K.V.M. acknowledges partial support via Lady Davis Postdoctoral Fellowship and Andrew and Erna Finci Viterbi Postdoctoral Fellowship.}
%\thanks{Parts of this work have appeared in 2016 Compressed Sensing Theory and its Applications to Radar, Sonar and Remote Sensing (CoSeRa) and 2017 IEEE International Radar Conference (RadarConf).}
}

% make the title area
\maketitle

\begin{abstract}
We present a cognitive prototype that demonstrates a colocated, frequency-division-multiplexed, multiple-input multiple-output (MIMO) radar which implements both temporal and spatial sub-Nyquist sampling. The signal is sampled and recovered via the Xampling framework. Cognition is due to the fact that the transmitter adapts its signal spectrum by emitting only those subbands that the receiver samples and processes. Real-time experiments demonstrate sub-Nyquist MIMO recovery of target scenes with $87.5$\% spatio-temporal bandwidth reduction and signal-to-noise-ratio of -10 dB.
%We present the design and hardware implementation of first prototype that demonstrates the principle of a sub-Nyquist colocated multiple-input multiple-output (MIMO) radar. The setup allows sampling in both spatial and temporal domains at rates much lower than dictated by the Nyquist sampling theorem. We use frequency division multiplexing (FDM) to achieve the orthogonality of MIMO waveforms and apply the Xampling framework for signal recovery. The prototype also implements a cognitive transmission scheme where each transmit waveform is restricted to those pre-determined subbands of the full signal bandwidth that the receiver samples and processes. Real-time experiments show reasonable recovery performance while operating as a $4\times5$ thinned random array wherein the combined spatial and spectral sampling factor reduction is $87.5$\% of that of a filled $8\times10$ array. 
\end{abstract}

\begin{IEEEkeywords}
MIMO radar, sub-Nyquist, compressed sensing, colocated array, cognitive radar, Xampling, prototype
\end{IEEEkeywords}

\IEEEpeerreviewmaketitle

\section{Introduction}
\label{sec:intro}
% MIMO radar: General
Multiple-input-multiple-output (MIMO) radar has been the topic of extensive research during the past decade \cite{fishler2004mimo,li2007mimo,bekkerman2006target}. This is largely because MIMO offers capabilities that outweigh an equivalent, standard phased array radar such as finer angular resolution \cite{boyer2011performance}, spatial diversity \cite{dianat2013target}, adaptive array realization \cite{huleihel2013optimal,xu2015joint}, and enhanced parameter identifiability \cite{haimovich2008mimo}. Similar to a standard phased-array radar, MIMO uses an array of several transmit (Tx) and receive (Rx) antenna elements. However, while a phased array transmits scaled versions of a single waveform, each of the MIMO radar transmitters may emit a different probing signal. The angular resolution of MIMO is equal to that of a \textit{virtual} uniform linear array (ULA) with the same antenna aperture but many more antenna elements.

% MIMO types
MIMO radars are usually classified as \textit{widely separated} or \textit{colocated} depending on the antenna placement. Widely separated MIMO antennas are located far from each other resulting in different radar cross sections (RCS) of the same target for each Tx-Rx antenna pair. This spatial diversity is advantageous in detection of a target with small backscatter and low speed \cite{dianat2013target,he2010target}. In a colocated MIMO radar \cite{li2007mimo,khan2014ambiguity}, the antenna elements are placed close to each other so that the RCS of a target appears identical to all the elements. The primary advantage of colocated MIMO is its high angular resolution \cite{godrich2010target} arising from its waveform diversity based on the mutual orthogonality - usually in time, frequency or code - of different transmit signals. A colocated MIMO receiver separates and coherently processes the target echoes corresponding to each transmitter in multiple Tx-Rx channels. In this paper, our focus is on colocated MIMO radar.

% Disadvantages of MIMO
When considering a MIMO radar, the spatial Nyquist theorem implies that an antenna array must not admit less than two signal samples per spatial period of the incident wave \cite{eldar2015sampling}. The spatial period is the operating wavelength $\lambda$ of the radar. Otherwise, \textit{spatial aliasing} is introduced leading to multiple beams in the antenna pattern, thereby reducing its directivity. In order to achieve high angular resolution, this implies a large virtual aperture with several elements that are located at least $\lambda/2$ spacing from each other. In sampling terms, the $\lambda/2$ ULA and the MIMO virtual ULA perform spatial sampling at the Nyquist rate. Even though a MIMO radar has less elements than an equivalent virtual phased array, it must employ multiple Tx-Rx chains resulting in huge hardware cost and very large computational complexity %associated with processing signals from multiple Tx-Rx chains in MIMO configurations is very large 
\cite{brookner2014mimo}. The range-time resolution of the radar is improved by transmitting signals with large bandwidth which necessitate large sampling rates %. Therefore, conventional processing resolutions of the radar in angular and range-time domains are limited by the number of antenna elements (or spatial sampling rate) and the receiver's temporal sampling rate, respectively. 
%Further, due to large transmit signal bandwidths, each receiver samples at high Nyquist rates 
leading to high energy consumption and additional cost of high-rate analog-to-digital converters (ADCs). When the MIMO transmit signals use waveform orthogonality based on frequency, the combined transmit spectrum may be excessively wider than conventional radars.

% CS methods in MIMO
Several methods have been proposed to address the problem of reducing the above-mentioned costs on hardware, energy and area in conventional MIMO radars (see e.g. \cite{ender2010compressive,mishra2018sub,cohen2018sub} for a review). Most exploit the fact that the target scene is \textit{sparse} facilitating the use of compressed sensing (CS) methods \cite{eldar2012compressed,eldar2015sampling}. Early works on CS-based MIMO radars focused on reducing only the temporal sampling rate in distributed \cite{gogineni2011target} and colocated configurations \cite{strohmer2013sparse,strohmer2014sparse}, including passive arrays \cite{malioutov2005sparse}. In \cite{kalogerias2014matrix,mishra2014compressed,sun2015mimomc}, the received signal samples from an array radar are processed as data matrices which, under certain conditions, are low rank. Random sampling in the temporal domain results in a partially observed data matrix whose missing entries may be retrieved using matrix completion methods. The target parameters are then recovered through classic Nyquist radar signal processing. All the above-mentioned applications of temporal CS still retain all antenna elements of a MIMO radar, thereby providing no reduction in the array hardware cost. Furthermore, while the receiver in these applications processes less measurements, the analog sampling with low-rate ADCs remains unexamined and CS-based target recovery procedures require dense sampling matrices.

Later works considered randomly reducing the number of antenna elements and employing CS-based target scene reconstruction \cite{carin2009relationship,yu2010mimo}. In \cite{rossi2014spatial}, spatial compressed sensing for random MIMO arrays was proposed and performance guarantees for recovery were provided. This setup was extended to phased array and phased-MIMO hybrids in \cite{mishra2017high}. Other sparse arrays such as spatial co-prime sampling for MIMO radars are suggested in \cite{qin2017doa}. A very recent work \cite{tohidi2018sparse} combines random reduction in both antenna elements and transmit pulses. Each receiver in these sparse MIMO array methods samples at the Nyquist rate and, therefore, requires high signal bandwidth and high-rate ADCs.

% SUMMeR
Recently, \cite{cohen2018summer} proposed a \textit{sub-Nyquist colocated MIMO radar} (or SUMMeR, hereafter) that can recover the target range and azimuth at native resolutions by simultaneously thinning an antenna array and sampling received signals at sub-Nyquist rates. The SUMMeR system was inspired by earlier works in \cite{baransky2014sub,barilan2014sub} which described sub-Nyquist sampling in a single antenna radar and demonstrated it through a custom-built prototype. The recovery algorithm relies on modeling the received radar signal with finite degrees of freedom per unit of time or as a finite-rate-of-innovation (FRI) signal \cite{eldar2015sampling}. The \textit{Xampling} framework is then used to obtain Fourier coefficients from the low-rate samples (or \textit{Xamples}) of this FRI signal \cite[p. 387-388]{eldar2015sampling}\cite{barilan2014sub}. Application of Xampling in space and time enables sub-Nyquist sampling without loss of any of the aforementioned radar resolutions. The Xamples are expressed as a matrix of unknown target parameters and the reconstruction algorithm is derived by extending orthogonal matching pursuit (OMP) \cite{eldar2015sampling} to simultaneously solve a system of CS matrix equations. A low-rate beamforming technique in the frequency domain called \textit{Doppler focusing} \cite{barilan2014sub} is added to the FRI-Xampling framework to also recover Doppler velocities along with delays and direction-of-arrival (DoA). %For an overview of other formulations of sub-Nyquist radars, we refer the reader to \cite{mishra2018sub}.

In SUMMeR, the radar antenna elements are randomly placed within the aperture, and signal orthogonality is achieved by frequency division multiplexing (FDM). In a conventional MIMO radar, the use of non-overlapping FDM waveforms results in strong range-azimuth coupling \cite{rabaste2013signal,xu2015joint} in the receiver processing, and therefore, it is common to use orthogonal code signals (i.e. code division multiplexing or CDM). The coupling due to FDM can be reduced by randomizing either the carrier frequencies across transmitters \cite{cohen2017high} or the element locations in the aperture. Our sub-Nyquist MIMO prototype adopts the latter approach. It employs narrow individual transmit bandwidth for high azimuth resolution and large overall total bandwidth for high range resolution. 

Here, we present a prototype that implements the SUMMeR concept in hardware by demonstrating sub-Nyquist sampling in both time and space in a real radar environment. %In this paper, we present the first hardware prototype of SUMMeR that is capable of demonstrating reduction in both spatial and spectral sampling using real-time analog signals. 
%We , developing a suitable hardware platform and finally reporting experimental results. 
Preliminary results of this work appeared in our conference publications \cite{mishra2016cognitive,cohen2017sub}. In this paper, we extend the prototype to implement a \textit{cognitive} MIMO radar. %and address real-time cognitive SUMMeR (CoSUMMeR) prototype challenges by evaluating the system requirements. 
%
%Additionally, our prototype implements a \textit{cognitive} MIMO radar. 
In recent years, cognitive radar has garnered considerable attention of the remote sensing community. The main advantage of such a system is its ability to learn the target environment and then adapt both the transmit and receive processing for optimal performance \cite{haykin2006cognitive,guerci2010cognitive}. Conventional radars can also optimize and change their processing techniques depending on the target scene, but their adaptability is restricted to receive processing only. Several possible radar cognition capabilities have been suggested where the environment specifications and corresponding suitable adaptive behaviors vary widely; examples include transmit beam-scheduling based on previous tracking history of the targets \cite{bell2015cognitive}, array adaptability and aperture sharing \cite{elbir2018cognitive}, and designing transmit waveform codes that avoid interfering bands by other licensed services \cite{aubry2015new,mishra2017performance,cohen2017spectrum,mishra2017auto,mishra2019toward}. 

In this work, we consider cognitive radar in the latter context of spectrum sharing which has also been recently explored between MIMO radar and MIMO communications \cite{li2016optimum,liu2018mu,qian2018joint}. Since the sub-Nyquist receiver samples and processes only a few disjoint subbands \cite{mishra2017performance,na2018tendsur}, our cognitive SUMMeR (CoSUMMeR) prototype transmits only in these subbands, leaving the rest of the transmit spectrum to be used by other services. %Our cognitive SUMMeR (CoSUMMeR) prototype enables spectrum sharing by restricting each transmit signal to only a few subbands leaving the rest of the transmit spectrum to be used by another service. Since our sub-Nyquist receiver samples and processes only a few disjoint subbands, we transmit in only these subbands to save the spectrum \cite{cohen2016towards,mishra2017performance,na2018tendsur}. 
Such a radar not only avoids radio-frequency (RF) interference from other licensed radiators in the vacant non-transmit subbands but also disguises the transmit frequencies as an effective electronic counter measure (ECM). Limiting the signal transmission to selective subbands allows for more in-band power resulting in an increase in signal-to-noise ratio (SNR). %Our prototype demonstrates this application of cognitive SUMMeR (CoSUMMeR). 

For a monostatic sub-Nyquist radar, \cite{baransky2014sub} presented the hardware realization of temporal sub-Nyquist sampling in radar through \textit{multiband} sampling \cite{mishra2017sub} in the receiver. Later, this prototype was extended to sub-Nyquist clutter removal in \cite{eldar2015clutter} and modified in \cite{cohen2016towards} to demonstrate monostatic cognitive sub-Nyquist radar. In these implementations, a few randomly chosen, narrow subbands of the received signal spectrum are pre-filtered before being sampled by low-rate ADCs. Since these implementations apply a bandpass filter and an ADC for each subband, a similar implementation of temporal sub-Nyquist sampling in each channel of a MIMO receiver would require enormous hardware resources. As we explain below in Section~\ref{sec:sysreq}, we circumvent such a design in CoSUMMeR by using the multiband version of foldable sampling as proposed in \cite{mishra2017sub,cohen2014channel}. This approach can be implemented by a single low-rate ADC to sample all subbands simultaneously and leads to realization of a compact, hand-held prototype. We mitigate the consequent increase in the subsampling noise through use of analog pre-processing filters with high stop-band attenuation. Our design strategy enables configuring the prototype either as a filled or thinned array, thereby allowing comparison of Nyquist and sub-Nyquist spatial sampling using the same hardware.

Hardware implementations of MIMO radars in surveillance applications are not common \cite{brookner2014mimo}%,brookner2015mimo}
. But, due to their ubiquitous illumination, MIMO radars have been very popular as both narrow and wide-band imaging sensors for far-field targets in synthetic aperture radar (SAR) \cite{ma2011three,wang2013mimo}, inverse SAR (ISAR) \cite{ma2012three} and ultrawideband imaging \cite{zhuge2011sparse}. Many MIMO imaging radar testbeds exist in X- \cite{klare2010mira,rommel2013orthogonal,belfiori2011tdma}, Ka- \cite{biallawons2013technical} and W-bands \cite{feger200977} for short-range applications such as environment monitoring, through-the-wall imaging, and automotive collision avoidance. The small size of antennas and other RF devices in these radar bands is helpful in building experimental testbeds. Very recently, miniature near-field MIMO imaging radar prototypes have been demonstrated \cite{pedross2017enhanced,fromenteze2017single}. Inspired by these implementations, we chose an X-band operating frequency for CoSUMMeR signal and target models. Our first real-time proof-of-concept experimental results show that, compared to a Nyquist $8$ Tx $\times$ $10$ Rx array, the CoSUMMeR array with $4$ Tx and $5$ Rx is capable of detecting targets up to SNR of -10 dB with spatial and temporal sampling rate reduction of $50\%$ and $80\%$, respectively. The CoSUMMeR design yields a combined spatio-temporal bandwidth reduction of $87.5$\% while also bringing down the number of hardware channels by $75$\%. Moreover, in low noise scenarios ($\sim-15$ dB), CoSUMMeR shows better detection performance than the Nyquist array.

In the next section, we %review state-of-the-art hardware implementations in MIMO and sub-Nyquist radars. In Section~\ref{sec:summer_theory}, we 
summarize the basic theory of our CoSUMMeR prototype. We evaluate the system requirements and formulate our design philosophy in Section~\ref{sec:sysreq}. A thorough description of the prototype is provided in Section~\ref{sec:sys_desc}, including all major sub-modules. We present results obtained by the prototype in real-time experiments in Section~\ref{sec:results} and conclude in Section~\ref{sec:summary}.

Throughout this paper, we reserve boldface lowercase, boldface uppercase and calligraphic letters for vectors, matrices and index sets, respectively. We denote the transpose and Hermitian by $(\cdot)^T$ and $(\cdot)^H$, respectively. The Kronecker %and Hadamard (point-wise) products are denoted by $\otimes$ and $\odot$, respectively. 
product is written as $\otimes$. The notation $\text{tr}\left\lbrace \cdot \right\rbrace $ is the trace of the matrix, $|\cdot|$ is the determinant and $\text{E}\left[ \cdot \right]$ is the statistical expectation function. %The notation $x \sim \mathcal{U}(u_l,u_u)$ represents a random variable drawn from the uniform distribution in the interval $[u_l,u_u]$ and $x \sim \mathcal{N}(\mu_x,\sigma_x^2)$ is the normal distribution with mean $\mu_x$ and variance $\sigma_x^2$. The functions $\text{max}$ and $\text{min}$ output the maximum and minimum value of their arguments, respectively. 
The function $\text{diag}(\cdot)$ outputs a diagonal matrix with the input vector along its main diagonal% while $\text{vec}(\cdot)$ vectorizes a matrix by stacking its columns
. The Fourier matrix $\mathbf{F}_N$ is a matrix of size $N\times N$ with $(n,k)$th entry given by $e^{\frac{-j2\pi nk}{N}}$. We use $\mathbf{I}_N$ for the identity matrix of size $N \times N$. Table~\ref{tbl:notations} summarizes some of the important notations in this paper.
%----------------------------------------------------------------
\begin{table}[t]	
%\captionsetup{labelfont={color=red},font={color=red}}
\caption{Glossary of notations}
        \label{tbl:notations}
		\vspace{-4mm}
		\footnotesize
        \begin{center}
		%{\color{red}
		\begin{tabular}{ |p{1.3cm}||p{6.7cm}|}
			\multicolumn{2}{c}{} \\
    		\hline
			Symbol & Description \\
			\hline		
			$T$ & Number of transmitters in a Nyquist MIMO array \\
            $R$ & Number of receivers in a Nyquist MIMO array \\
			$M$ & Number of transmitters in a sub-Nyquist MIMO array \\
			$Q$ & Number of receivers in a sub-Nyquist MIMO array\\
			$Z$ & Normalized aperture\\
			$\xi_m$ & Location of $m$th transmitter in SUMMeR array\\
			$\zeta_q$ & Location of $q$th receiver in SUMMeR array\\
			$Z$ & Normalized aperture\\
            $\lambda$ & Operating wavelength \\
            $f_c$ & Common carrier frequency \\
            $c$ & Speed of light \\
            $h_m(t)$ & Non-cognitive time-domain pulse of $m$th transmitter\\
            $\widetilde{h}_m(t)$ & Cognitive time-domain pulse of $m$th transmitter\\
			$H_m(\omega)$ & CTFT of $h_m(t)$\\
			$\widetilde{H}_m(\omega)$ & CTFT of $\widetilde{h}_m(t)$\\
			$f_m$ & Center frequency of $h_m(t)$\\
			$T_p$ & Pulse width\\
            $\tau$ & PRI \\
            $P_t$ & Total transmit power at each transmitter \\
			$L$ & Number of targets \\
			$R_l$ & Range of the $l$th target\\
			$\tau_l$ ($\hat{\tau}_l$) & True (estimated) time delays of the $l$th target\\
            $\alpha_l$ & Reflectivity of the $l$th target \\
			$v_l$ & Doppler velocity of the $l$th target \\
			$f_l^D$ ($\hat{f}_l^D$) & True (estimated) Doppler frequency of the $l$th target \\
			$\theta_l$ & Azimuth angle of the $l$th target \\
			$\vartheta_l$ ($\hat{\vartheta}_l$) & True (estimated) Azimuth parameter $\sin\theta_l$\\ $\beta_{mq}$ & Array structure parameter\\
            $x_q(t)$ & Baseband receive signal at $q$th receiver\\
            $x_q^p(t)$ & Baseband receive signal at $q$th receiver for $p$th pulse\\
            $y_{m,q}^{p}$ & Fourier coefficients for $m$th channel, $q$th receiver, $p$th pulse \\
            $\Phi^{\nu}_{m,q}$ & Fourier coefficients $y_{m,q}^{p}$ after focusing at frequency $\nu$\\
            $N$ & Number of Fourier coefficients in each channel\\
            $\kappa$ & Set of sampled Fourier coefficients per channel\\
			$K$ & Number of sampled Fourier coefficients per channel \\
			$\bm{Z}^{m}$ & All sampled Fourier coefficients for $m$th transmitter\\
            $\bm{A}^{m}$ & Range dictionary \\
            $\bm{B}^{m}$ & Azimuth dictionary \\
            $\bm{F}_{P}$ & Fourier matrix of size $P\times P$ \\
            $\bm{X}_{D}$ & Sparse matrix with non-zero elements at target locations\\
            $B_h$ & Non-cognitive bandwidth of each transmitter\\
            $N_b$ & Number of sub-bands in each cognitive transmit pulse\\
            $\mathcal{B}_i$ & Set of frequencies in $i$th subband\\
            \hline
		\end{tabular}%	
        %}
        \end{center}
		%
		%	\vspace*{-5mm}
\end{table}		  
%----------------------------------------------------------------
\section{Basic Theory of CoSUMMeR}
\label{sec:summer_theory}
\begin{figure}
\includegraphics[width=0.45\textwidth]{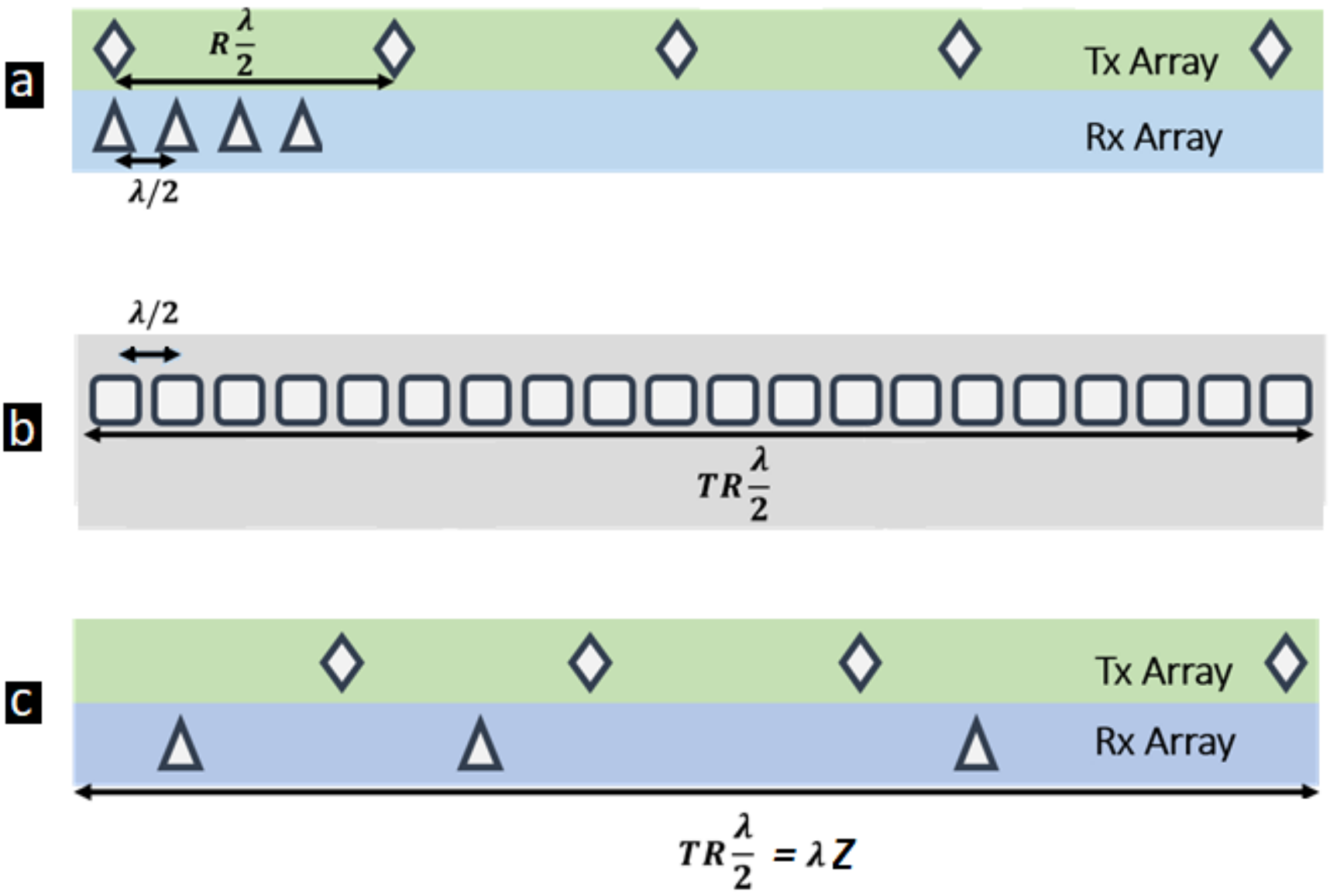}
\caption{Location of transmit (diamonds) and receive (triangles) antenna elements within the same physical aperture for (a) conventional MIMO array with $T=5$ transmitters and $R=4$ receivers, (b) virtual ULA with $TR=20$ antenna elements, and (c) randomly thinned MIMO array with $M=4$ transmitters and $Q=3$ receivers.}
\label{fig:all_arrays}
\end{figure}

Except for the cognitive transmission, our CoSUMMeR prototype follows the signal model and algorithms suggested in \cite{cohen2018summer}, and hence we only summarize them here. Let the operating wavelength of the radar be $\lambda$ and the total number of transmit and receive elements in a standard colocated MIMO radar array be $T$ and $R$, respectively. The classic MIMO approach adopts a virtual ULA structure, where the receive antennas spaced by $\frac{\lambda}{2}$ and transmit antennas spaced by $R\frac{\lambda }{2}$ form two ULAs (or vice versa). Here, the coherent processing of a total of $TR$ channels in the receiver creates a virtual equivalent of a phased array with $TR$ $\frac{\lambda }{2}$-spaced receivers and normalized aperture $Z=\frac{TR}{2}$. This standard array structure and the corresponding receiver virtual array are illustrated in Fig.~\ref{fig:all_arrays}a-b for $T=5$ and $R=4$.
%---------------------------------------------------------------------------------
%---------------------------------------------------------------------------------

Consider now a sub-Nyquist colocated MIMO radar system that has $M<T$ transmit and $Q<R$ receive antennas. The locations of these antennas are chosen uniformly at random within the aperture of the virtual array mentioned above, that is $\{\xi_m\}_{m=0}^{M−1} \sim \mathcal{U}[0, Z]$ and $\{\zeta_q\}_{q=0}^{Q−1} \sim \mathcal{U}[0, Z]$, respectively. An example can be seen in Fig.~\ref{fig:all_arrays}c with $M=4$ and $Q=3$.
The $m$th transmitting antenna sends $P$ pulses ${{s}_{m}}\left( t \right)$ given by
\begin{equation}
%\label{eq:trMth}
    \begin{array}{lll}
     {{s}_{m}}\left( t \right)=\sum\limits_{p=0}^{P-1}{{{h}_{m}}\left( t - p\tau\right)}{{e}^{j2\pi {{f}_{c}}t}},\quad0\le t\le P\tau,
    \end{array}
\end{equation}
where $\tau$ denotes the pulse repetition interval (PRI), $P\tau$ is the coherent processing interval (CPI), ${{f}_{c}} = c/\lambda$ is the common carrier frequency at the radio frequency (RF) stage, $c$ is the speed of light, and $\{{{h}_{m}}\left( t \right)\}_{m=0}^{M-1}$ is a set of narrowband, orthogonal FDM pulses centered at the modulating frequency $f_m$ each with continuous-time Fourier transform (CTFT)
\begin{equation}
%\label{eq:ctft}
H_m(\omega) = \int\limits_{-\infty}^{\infty}h_m(t)e^{-j\omega t}dt.
\end{equation}
For simplicity, we assume that $f_c \tau$ is an integer, so that the delay $e^{-j2\pi f_c \tau p}$ is canceled in the modulation. The pulse time support is denoted by $T_p$.

Assume a target scene consisting of $L$ non-fluctuating point targets following the Swerling-0 model \cite{skolnik2008radar} whose locations are given by their ranges $R_l$, Doppler velocity $v_l$, and azimuth angles $\theta_l$, $1 \le l \le L$. The pulses transmitted by the radar are reflected back by the targets and collected at the receive antennas. When the received waveform is downconverted from RF to baseband, we obtain the following signal at the $q$th antenna,
\begin{equation}
x_q \left( t \right) = \sum\limits_{p=0}^{P-1}\sum\limits_{m=0}^{M-1} \sum\limits_{l=1}^{L} \alpha_l h_m \left( t-p\tau-\tau _{l} \right) e^{j2 \pi \beta_{mq} \vartheta _l}e^{j 2\pi f^D_l p \tau},
\end{equation}
where $\alpha_l$ denotes the complex-valued reflectivity of the $l$th target, $\tau_l = 2R_l/c$ is the range-time delay the $l$th target, $f_l^D = \frac{2v_l}{c}f_c$ is the frequency in the Doppler spectrum, $\vartheta_l = \sin\theta_l$ is the azimuth parameter, and $\beta_{mq} = (\zeta_q + \xi_m)(f_m\frac{\lambda}{c}+1)$ is governed by the array structure. For processing, we express $x_q(t)$ as a sum of single frames
\begin{equation}
\label{eq:frames}
x_q(t)= \sum_{p=0}^{P-1} x_q^p(t),
\end{equation}
where
\begin{equation}
\label{eq:one_frame}
x_q^p(t)= \sum_{m=0}^{M-1} \sum_{l=1}^{L} \alpha_l h(t-\tau_l - p\tau) e^{j2 \pi \beta_{mq} \vartheta _l} e^{j 2\pi f^D_l p \tau}.
\end{equation}
The radar goal is to estimate the time delay ${{\tau}_{l}}$, azimuth ${{\theta }_{l}}$, and Doppler shifts $f_l^D$ of each target from low rate samples of $x_q(t)$, and a small number of $M$ channels and $Q$ antennas.

\subsection{Xampling in Time and Space}
\label{subsec:xampling}
The application of Xampling in both space and time enables recovery of range, direction and velocity at sub-Nyquist rates \cite{cohen2018summer}. The sampling technique is the same as in temporal sub-Nyquist radar \cite{barilan2014sub}, except that now the low-rate samples are obtained in both range and azimuth domains. 

The received signal $x_q(t)$ is separated into $M$ channels, aligned, and normalized. The Fourier coefficients of the received signal corresponding to the channel that processes the $m$th transmitter echo at the $q$th receiver are 
\begin{equation}
\label{eq:coffAlligned}
   y_{m,q}^p[k]=   \sum_{l=1}^{L} \alpha_l e^{j2\pi \beta_{mq} \vartheta_l}e^{-j\frac{2\pi }{\tau}k\tau_l} e^{-j2\pi f_m \tau_l} e^{j 2 \pi f^D_l p \tau},
\end{equation}
where $-\frac{N}{2} \leq k \leq -\frac{N}{2}-1$, $f_m$ is the (baseband) carrier frequency of the $m$th transmitter and $N$ is the number of Fourier coefficients per channel. Xampling obtains a set $\kappa$ of arbitrarily chosen Fourier coefficients from low rate samples of the received channel signal such that $|\kappa| = K < N$.

As in traditional MIMO, assume that the time delays, azimuths and Doppler frequencies are aligned to a grid. In particular, $\tau_l = \frac{\tau}{TN}s_l$, $\vartheta_l = -1+\frac{2}{TR} r_l$, and $f^D_l = -\frac{1}{2\tau}+\frac{1}{P\tau}u_l$, where $s_l$, $r_l$, and $u_l$ are integers satisfying $0 \leq s_l \leq TN-1$, $0 \leq r_l \leq TR -1$, and $0 \leq u_l \leq P-1$, respectively.

Let $\mathbf{Z}^m$ be the $KQ \times P$ matrix with $q$th column given by the vertical concatenation of $y_{m,q}^p[k], k \in \kappa$, for $0 \leq q \leq Q-1$. We can then write $\mathbf{Z}^m$ as
\begin{equation}
\label{eq:model2}
\mathbf{Z}^m = \left( \mathbf{{B}}^m \otimes  \mathbf{A}^m \right) \mathbf{X}_D \mathbf{F}^H_P.
\end{equation}
Here, $\mathbf{A}^m$ denotes the $K \times TN$ matrix whose $(k,n)$th element is $e^{- j \frac{2 \pi}{TN} \kappa_kn} e^{-j2\pi \frac{f_m}{B_h} \frac{n}{T}}$ with $\kappa_k$ the $k$th element in $\kappa$, $\mathbf{B}^m$ is the $Q \times TR$ matrix with $(q,p)$th element $e^{-j2 \pi \beta_{mq} (-1 +\frac{2}{TR}p)}$ and $\mathbf{F}_P$ denotes the $ P \times P$ Fourier matrix. The matrix $\mathbf{X}_D$ is a $T^2NR \times P$ sparse matrix that contains the values $\alpha_l$ at the $L$ indices $\left( r_l TN +s_l, u_l \right)$. The temporal, spatial and frequency resolutions stipulated by $\mathbf{X}_D$ are $\frac{1}{TB_h}$, $\frac{2}{TR}$, and $\frac{1}{P \tau}$ respectively. The range and azimuth dictionaries $\mathbf{A}^m$ and $\mathbf{B}^m$ are not square matrices due to the low-rate sampling of Fourier coefficients at each receiver and reduction in antenna elements, respectively. Therefore, the system of equations in (\ref{eq:model2}) is undetermined in azimuth and range. %Our goal is to recover $\mathbf{X}_D$ from the measurement matrices $\mathbf{Z}^m, 0 \leq m \leq M-1$.

The minimum required number of SUMMeR transmit and receive elements as well as samples $K$ to recover $\mathbf{X}_D$ depend only on the number of targets present as stated by Theorem~\ref{th:cond2} below. These design resources are, therefore, substantially fewer than the requirements of a Nyquist MIMO array. 
\begin{theorem}\cite{cohen2018summer}
\label{th:cond2}
The minimal number of transmit and receive array elements, i.e. M and Q, respectively, required for perfect recovery of $\mathbf{X}_D$ with $L$ targets in a noiseless setting are determined by $MQ \geq 2L$. In addition, the number of samples per receiver is at least $MK \geq 2L$ where $K$ is the number of Fourier coefficients sampled per receiver and the number of pulses per transmitter is $P \geq 2L$.
\end{theorem}

\subsection{Range-Azimuth-Doppler Recovery}
To jointly recover the range, azimuth and Doppler frequencies of the targets, Doppler focusing \cite{barilan2014sub} is used in SUMMeR which, for a specific frequency $\nu$, yields
\begin{eqnarray}
\label{eq:dop_reduced}
   \Phi^{\nu}_{m,q}[k] &=&\sum_{p=0}^{P-1} y_{m,q}^p[k] e^{-j2\pi \nu p \tau} \\
 &=&   \sum_{l=1}^{L} \alpha_l e^{j2\pi \beta_{mq} \vartheta_l}e^{-j\frac{2\pi }{\tau}(k+f_m \tau)\tau_l} \sum_{p=0}^{P-1} e^{j 2 \pi (f^D_l -\nu) p\tau}, \nonumber
\end{eqnarray} 
for $-\frac{N}{2} \leq k \leq -\frac{N}{2}-1$. Since 
\begin{equation}
\label{eq:dop_foc}
\sum_{p=0}^{P-1} e^{j 2 \pi (f^D_l -\nu) p\tau} \cong \begin{dcases} P & |f^D_l -\nu| < \frac{1}{2P\tau}, \\
0 & \text{otherwise}, \end{dcases}
\end{equation} 
for each focused frequency $\nu$,  (\ref{eq:dop_reduced}) reduces to a 2-D problem, which can be solved using CS recovery techniques. %, as summarized in Algorithm \ref{algo:omp2}.
%We note that Doppler focusing increases the SNR by a factor a $P$, as can be seen in (\ref{eq:dop_foc}).
We refer the reader to \cite{cohen2018summer} for full details of this recovery algorithm. Note that Doppler focusing can be efficiently performed using the fast Fourier transform (FFT). %We point out that since Doppler focusing and the projection step (step 2 in Algorithm \ref{algo:omp}) are both linear time-invariant (LTI), they can be interchanged. In Algorithm \ref{algo:omp2}, we choose to perform the projections before Doppler focusing, for simplicity. 
%To that end, we reformulate (\ref{eq:model2}) as follows
%\begin{equation}
%\mathbf{Y}^{(m,p)} = \mathbf{A}^m \mathbf{X}^p \mathbf{B}^m,
%\end{equation}
%for $0 \leq m \leq M-1$ and $0 \leq p \leq P-1$. In fact, $\mathbf{Y}^{(m,p)}$ reorganizes the $p$th column of $\mathbf{Z}^m$ into a $K \times Q$ matrix and $\mathbf{X}^p$ reshapes the $p$ column of $\mathbf{X}_D$ into a $TN \times TR$ matrix. For simplicity, we choose $P=P_0$.
%In the algorithm description, $\text{vec}(\mathbf{Z})$ concatenates the columns of $\mathbf{Z}^m$, $\mathbf{e}_t(l)= \left[ (\mathbf{e}_t^0(l))^T \, \cdots \, (\mathbf{e}_t^{M-1}(l))^T \right]^T$, where $\mathbf{e}_t^m(l) = \text{vec}( (\mathbf{\bar{B}}^m \otimes \mathbf{A}^m)_{\Lambda_t(l,2)TN+\Lambda_t(l,1)} \left((\mathbf{\bar{F}}^m)^T_{\Lambda_t(l,3)} \right)^T)$ with $\Lambda_t(l,i)$ the $(l,i)$th element in the index set $\Lambda_t$ at the $t$th iteration, and $\mathbf{E}_t= [\mathbf{e}_t(1) \, \dots \, \mathbf{e}_t(t)]$.
Once $\mathbf{X}_D$ is recovered, the delays, azimuths and Dopplers are estimated as
\begin{equation}
\hat{\tau}_l = \frac{\tau \Lambda_L(l,1)}{TN}, \, \hat{\vartheta}_l = -1+\frac{2 \Lambda_L(l,2)}{TR}, \, \hat{f}_l^D = -\frac{1}{2\tau}+\frac{\Lambda_L(l,3)}{P\tau},
\end{equation}
where $\Lambda_t(l,i)$ is the $(l,i)$th element in the index set $\Lambda_t$ at the $t$th iteration of the recovery algorithm. Since in real scenarios, target delays, Dopplers and azimuths are not necessarily aligned to a grid, a finer grid can be used around detection points on the coarse grid to reduce quantization error.

\subsection{CoSUMMeR}
\label{subsec:cognitive}
%-----------------------------------------------------------------------------------
\begin{figure}
  \includegraphics[scale=0.25]{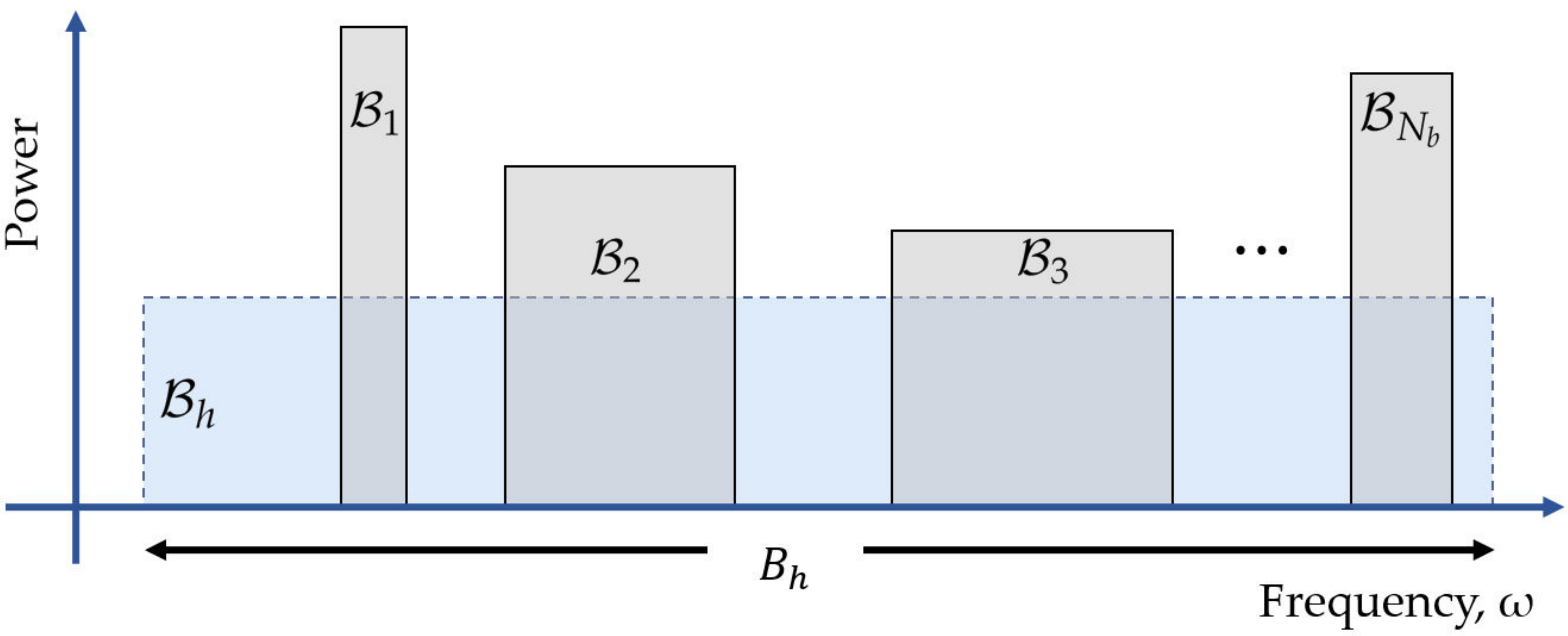}
  \caption{Spectrum of a single transmit signal of a conventional radar uses the full bandwidth $\mathcal{B}_h$. A cognitive radar transmits only in subbands $\{\mathcal{B}_i\}_{i=1}^{N_b}$, but with more in-band power than the conventional radar.}
	\label{fig:cogspec}
\end{figure}
%-----------------------------------------------------------------------------------
%-----------------------------------------------------------------------------------
\begin{figure}[!t]
\centering
%\subfloat[FDM Nyquist MIMO]{%
%  \includegraphics[width=0.32\textwidth]{fdm_nyquist_spec.pdf}
%  \label{fig:fdm_nyq_spec}%
%}\qquad
%\subfloat[SUMMeR]{%
%  \includegraphics[width=0.32\textwidth]{summer_spec.pdf}
%  \label{fig:summer_spec}%  
%}\qquad
%\subfloat[CoSUMMeR]{%
%  \includegraphics[width=0.32\textwidth]{cosummer_spec.pdf}
%  \label{fig:cosummer_spec}%  
%}
\includegraphics[width=1.0\columnwidth]{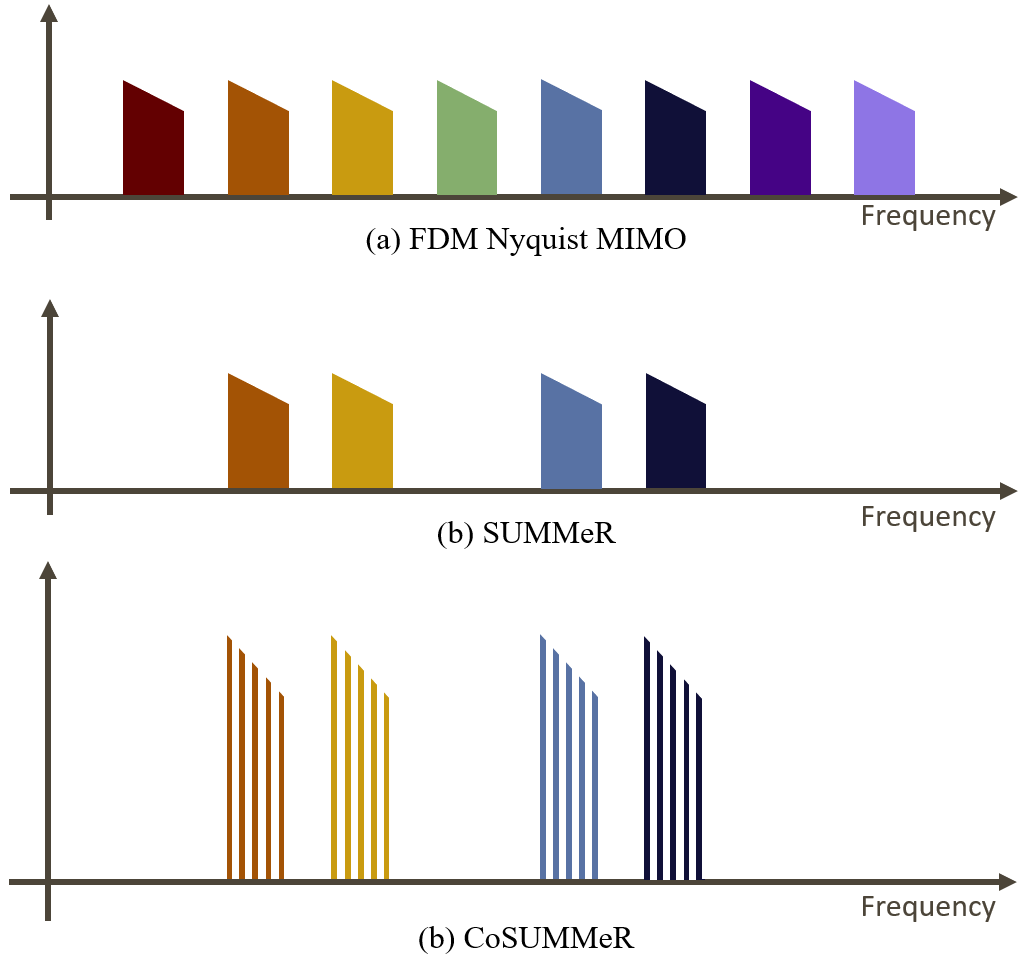}
\caption{An illustration of the one-sided transmit spectrum of (a) FDM-based conventional (spatial Nyquist) MIMO radar employing $T=8$ transmitters, (b) SUMMeR with $M=4$ transmitters, and (c) CoSUMMeR with each transmit signal restricted to only a few subbands but with more in-band power than SUMMeR.}
\label{fig:spec_comp}    
\end{figure}
%-----------------------------------------------------------------------------------
%-----------------------------------------------------------------------------------
\begin{figure*}[!t]
\centering
  \includegraphics[width=0.8\textwidth]{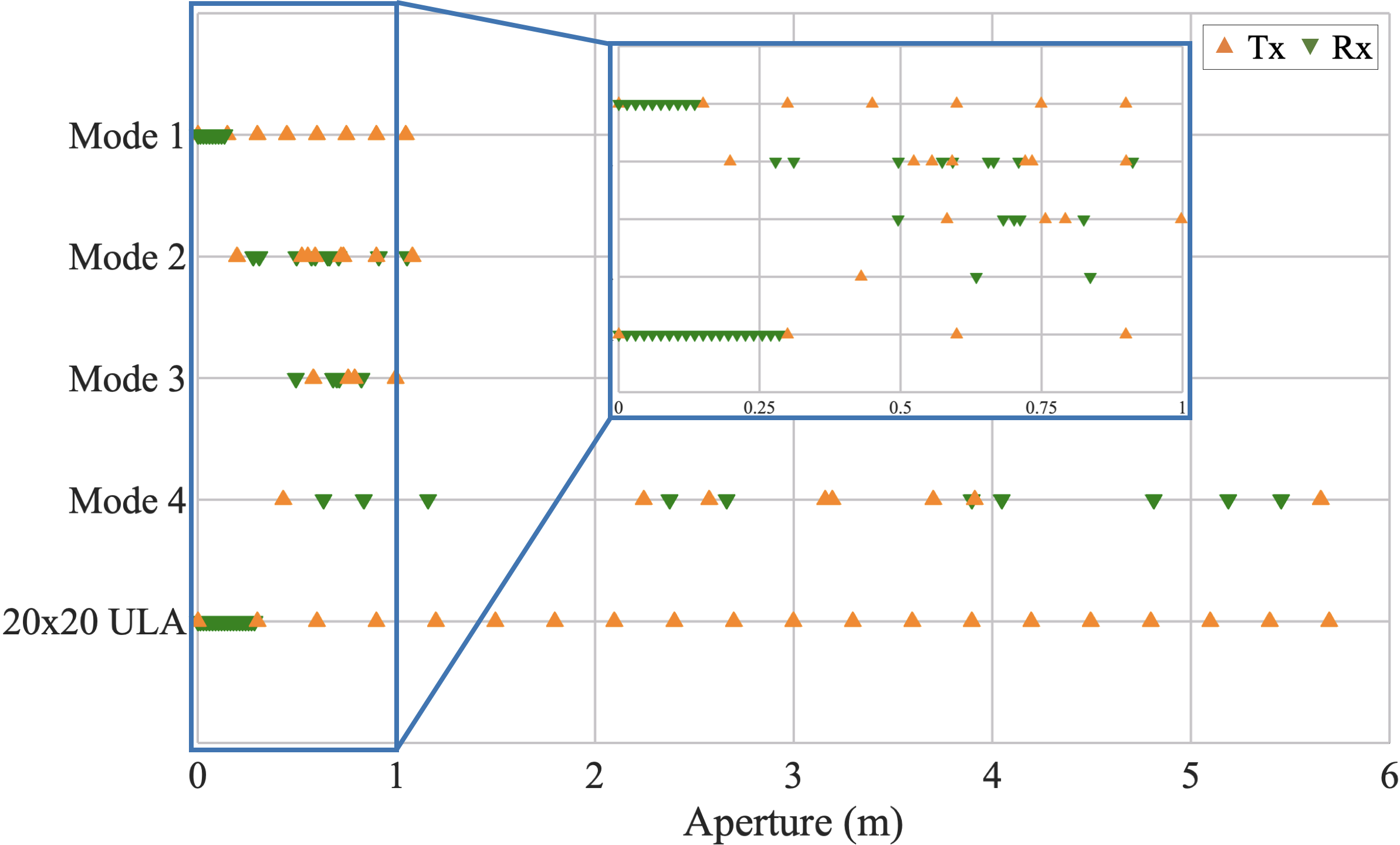}
  \caption{Tx and Rx element locations for the hardware prototype modes over a 6 m antenna aperture. Mode 4's virtual array equivalent is the $20\times20$ ULA. The inset plot shows the selected region on a magnified scale.}
\label{fig:arrays}
\end{figure*}
%-----------------------------------------------------------------------------------
So far, we focused on processing the received signal in sub-Nyquist MIMO radar. The receiver design in the sub-Nyquist framework can be exploited to also alter the behavior of the radar transmitter. Let us assume that $\mathcal{B}_h$ is the set of all frequencies in the single transmitter signal spectrum of effective bandwidth $B_h$. In cognitive radar transmission, the spectrum $\widetilde{H}_m(\omega)$ of each of the cognitive transmitted waveforms $\{\widetilde{h}_m(t)\}_{m=1}^{M}$ is limited to a total of $N_b$ non-overlapping frequency bands $\mathcal{B}_i$, $1 \le i \le N_b$ (see Fig.~\ref{fig:cogspec}):
\begin{align}
\widetilde{H}_m(\omega)
&= \begin{dcases} 
    \gamma(\omega) H_m(\omega), \phantom{1}\phantom{1} \omega \in \bigcup_{i=1}^{N_b}\mathcal{B}_i \subset \mathcal{B}_h\\
    0, \phantom{1}\phantom{1}\text{otherwise},
   \end{dcases}
\end{align}
where $\gamma(\omega)={B_h}/{|\mathcal{B}_i}|$ for $\omega \in \mathcal{B}_i$. The total transmit power $P_t$ remains the same such that the power relation between the conventional and cognitive waveforms is
\begin{align}
\int_{-B_h/2}^{B_h/2} |H_m(\omega)|^2\, \mathrm{d}\omega = \sum_{i=1}^{N_b}\int\limits_{\mathcal{B}_i} |\widetilde{H}_m(\omega)|^2\, \mathrm{d}\omega = P_t.
\end{align}
In a cognitive radar, the sub-Nyquist receiver obtains the set $\kappa$ of the Fourier coefficients only from the subbands $\mathcal{B}_i$. Note that a conventional radar that employs a Nyquist receiver will be unable to process echoes from disjoint subbands.

Cognitive transmission imparts two advantages to the CoSUMMeR hardware. First, as we explain in Section~\ref{subsec:temp_sn}, the spatial sub-Nyquist processing of large arrays can be easily designed without replicating the pre-filtering operation for each subband in the hardware. Second, since the total transmit power remains the same, a cognitive signal has more in-band power resulting in an increase in SNR. For a monostatic cognitive sub-Nyquist radar, our earlier work \cite{mishra2017performance} compares the performance of conventional and cognitive radars using the extended Ziv-Zakai lower bound (EZB) for the particular case of time delay estimation of a single target. %In a conventional radar, the EZB for a single target delay estimate $\hat{\tau_0}$ is 
%\begin{align}
%\label{eq:ezzlb_con}
%EZB_R(\hat{\tau_0}) = \sigma^2_{\tau_0}\cdot 2Q\left(\sqrt{\dfrac{SNR}{2}}\right) + \dfrac{\Gamma_{3/2}\left(\dfrac{SNR}{4}\right)}{SNR\cdot\overline{F}^2},
%\end{align}
%where $Q(\cdot)$ denotes the right tail Gaussian probability function, $\sigma^2_{\tau_0}$ is the prior variance of delay $\tau_0$, $\Gamma_{a}(b)$ is the incomplete gamma function with parameter $a$ and upper limit $b$, and $\overline{F}$ is the root-mean-square (rms) bandwidth of the full-band signal. The bound for cognitive radar is given in the following theorem.
%\begin{theorem}\cite{mishra2017performance}
%\label{thm:ezb_crr}
%The extended Ziv-Zakai lower bound (EZB) for delay estimation in a cognitive radar is 
%\begin{flalign}
%\label{eq:ezzlb_cog}
%EZB_{CR}(\hat{\tau}_0) &= \sigma^2_{\tau_0}\cdot 2Q\left(\sqrt{\dfrac{\widetilde{SNR}} {2}}\right) + \dfrac{\Gamma_{3/2}\left(\dfrac{\widetilde{SNR}}{4}\right)}{\sum\limits_{i=1}^{N_b}SNR_i\cdot\overline{F_i^2}},
%\end{flalign}
%where $SNR_i$ and $\overline{F_i}$ are the in-band SNR and rms bandwidth of the $i$th subband and $\widetilde{SNR}$ is the total SNR. 
%\end{theorem}
As noted in \cite{mishra2017performance}, %since $\sum\limits_{i=1}^{N_b}\mathcal{B}_i \subset B_h$, we have $\widetilde{SNR} > SNR$ for given $P_t$. Therefore, 
the SNR threshold for asymptotic performance of EZB in cognitive radar is lower than that of a conventional radar. When the noise increases and power remains constant for both radars, the asymptotic performance of cognitive radar EZB is more tolerant to noise. In this context, experimental results presented later in Section~\ref{sec:results} provide a proof-of-concept of the theoretical intuition that CoSUMMeR performs better than non-cognitive Nyquist and sub-Nyquist MIMO radars in low SNR scenarios.

Since the waveform orthogonality of transmit waveforms in SUMMeR is based on FDM, spatial compression through random removal of antenna elements in the transmit array also eliminates spectrum usage by those transmitters, as we illustrate in Fig.~\ref{fig:spec_comp}. The spectrum savings are further improved in CoSUMMeR through use of a small portion of the available bandwidth of the SUMMeR signal; the range resolution is not lost while sampling at a low rate.

In a spectrally crowded operational environment, the radar chooses its subband locations based on the interference levels at all available frequencies in the spectrum. In practice, a radar could obtain the spectral occupancy information through passive (receive-only) sweeping across the desired spectrum. Spectral coexistence comprises some exchange of information between the radar and a coexisting communications system where the latter conveys the location of its signals to the radar. During the cognitive operation, equipped with the information from the communications system and passive sensing, the objective of the radar is to identify an appropriate transmit frequency set such that the radar's probability of detection $P_d$ is maximized. For a fixed probability of false alarm $P_{\text{fa}}$, the $P_d$ increases with higher signal-to-interference-and-noise-ratio (SINR). Hence, the frequency selection criterion is to choose subbands so that the SINR is maximized at the receiver. Our previous work \cite{cohen2017spectrum} maps the cognitive subband selection problem to seeking a block-sparse vector and solves it by employing a structured greedy algorithm. This process is repeated every few scans.

\section{Design Philosophy}
\label{sec:sysreq}
We now discuss the various design considerations for our CoSUMMeR prototype with the goal of designing a compact, portable system that demonstrates theoretical concepts of CoSUMMeR in practice. 

\subsection{Experimental Environment}
\label{subsec:exp_env}
The spatial and temporal sampling aspects of the sub-Nyquist MIMO prototype manifest only in the receiver processing. Therefore, we do not physically radiate the transmit waveforms from an antenna. Instead, we employ National Instruments AWR software design environment which is capable of simulating complex sensing scenes including multiple targets, their ranges, Doppler velocities, RCS, and propagation losses in the medium. The AWR, therefore, provides reflected signals from the targets as received at the antenna waveguide front and is capable of modeling an RF receiver response with realistic component-level simulations. It outputs the demodulated signal at the intermediate-frequency (IF) stage of the analog receiver. We record the AWR output of the received signal at baseband for multiple transmit waveforms and target scenarios. The complex samples (in-phase $\mathcal{I}$ and quadrature-phase $\mathcal{Q}$ pairs) of the received signal are then stored in an on-board memory of a custom-designed waveform generator board. The prototype processes these pre-recorded signals in real-time. We omit the implementation of the up-conversion to RF carrier frequency in the transmitter and the corresponding down-conversion in the receiver from this prototype. For reasons explained in Section~\ref{sec:intro}, we assume that the physical array aperture and simulated target response correspond to an X-band ($f_c = 10$ GHz) radar. %The choice of radar frequency band also affects the clutter response that we intend to consider in a future extension of this prototype.

\subsection{Spatial Sub-Nyquist Sampling}
\label{subsec:spat_sn}
We consider the implementation of a MIMO radar architecture with resources that can handle 8 Tx and 10 Rx antenna elements. These elements can be positioned within a given physical aperture in several different constellations or \textit{modes}. While operating at its maximum strength of 8 Tx and 10 Rx elements, the prototype can be programmed as a standard Nyquist array or ULA (Mode 1) with an equivalent aperture of an $8\times10$ virtual array at X-band, i.e. $1.2$ m. In Mode 2, the prototype utilizes the same $8\times10$ elements as a random array within the physical aperture of $1.2$ m. Note that the equivalent virtual aperture of this array may be wider than the Nyquist Mode 1. 

For the sub-Nyquist array, the prototype offers two options. It can operate as a thinned $4\times5$ array (Mode 3) with the corresponding Nyquist array being Mode 1. In this constellation, the receive channels corresponding to the removed transmit elements are not processed by the digital receiver. Mode 3, hence, demonstrates spatial sub-Nyquist sampling. The prototype can process the received signals for the same target scenario in both Mode 1 and 3 to allow for a comparison between spatial Nyquist and sub-Nyquist sampling methods.

In the second spatial sub-Nyquist configuration, the prototype functions as a $8\times10$ thinned array (Mode 4) for which the equivalent Nyquist antenna is the virtual $20\times20$ array with an aperture of $6$ m. However, in this case, with a resource limit of $8$ Tx and $10$ Rx, the prototype is unable to compare the Nyquist and sub-Nyquist arrays. Nonetheless, this provides an opportunity to evaluate sub-Nyquist processing at higher angular resolution than all other three modes. Figure~\ref{fig:arrays} shows exact details of element locations for all four modes.

In a random array, antenna selection may be decided through advanced methods which optimize element placement for improved parameter estimation. These approaches include greedy search \cite{tohidi2018sparse} or deep learning \cite{elbir2018cognitive}. The random antenna placement in our prototype was determined through software simulations which performed extensive search for optimum arrays. The random arrays that provided localization with minimum error for multiple targets were selected.

\subsection{Temporal Sub-Nyquist Sampling}
\label{subsec:temp_sn}
A conventional 8x10 MIMO radar receiver would require simultaneous hardware processing of 80 complex data streams (or 160 $\mathcal{I}$ and $\mathcal{Q}$ channels). A separate sub-Nyquist receiver for each of these 80 channels is expensive and increases the hardware size. Hence, we implement the eight channel analog processing chain for only one receive antenna element and then serialize the received signals of all 10 elements through this chain. The analog processing chain consists of eight different filtering channels which separate the received signal corresponding to each of the $8$ transmitters. This approach allows the prototype to implement a number of receivers even greater than 10 because the eight-channel hardware places an upper limit on only the number of transmitters.

Given a particular receive element, we need to extract a set of Fourier coefficients $\kappa$, $|\kappa| = K$, (see Section~\ref{subsec:xampling}) from low rate samples of each of its transmit channels. It has been shown \cite[pp. 210-268]{eldar2012compressed} that high recovery performance is attained when these coefficients are drawn uniformly at random. An ADC cannot, however, individually acquire each of the randomly chosen Fourier coefficients conveniently. Several practical implementations of sub-Nyquist receivers exist including the modulated wideband converter (MWC) \cite{mishali2010theory}. In particular, the sub-Nyquist radar prototype in \cite{baransky2014sub} opted for \textit{multiband sampling} where random disjoint subsets of $\kappa$ are sampled, with each subset containing consecutive Fourier coefficients. This prototype used four random Fourier coefficient groups, pre-filtered the baseband signal to corresponding four subbands (or Xampling \textit{slices}), and sampled each subband through a separate low-rate ADC. 

If we use the same pre-filtering approach as in \cite{baransky2014sub} for each of the eight channels of our sub-Nyquist MIMO prototype, then the hardware design would require a total of $4\times8=32$ bandpass filters (BPFs) and ADCs excluding the analog filters to separate transmit channels. Moreover, the hardware cost would increase dramatically if the number of subbands are increased or separate analog channels are implemented for each receiver. We sidestep this requirement by adopting cognitive transmission wherein the analog signal of each channel lives only in certain pre-determined subbands and consequently, a BPF stage for each subband is not required. 

More importantly, for each channel, a single low-rate ADC can \textit{subsample} this narrow-band signal as long as the subbands are \textit{coset} bands so that they do not alias after sampling. This is based on \textit{foldable sampling} suggested for ultra-wideband reception in \cite{cohen2014channel} where the received signal is sampled below the Nyquist rate resulting in folding over or aliasing of the signal spectrum that exceeds the Nyquist sampling rate $f_s$ of the ADC. The signal distortion due to aliasing is avoided by notching out those Fourier coefficients from the transmit signal that will alias over the low-frequency part of the sampled signal. In particular, assume that the entire received signal consisting of $N$ Fourier coefficients is divided into $q$ groups of $K$ coefficients. If we subsample this signal at the rate $f_s/q$, then $K$ coefficients of each group will alias over each other in the sampled signal. In \cite{cohen2014channel}, $K$ Fourier coefficients are chosen randomly from any of the $q$ groups and the remaining $q-1$ aliasing coefficients are notched out in the transmit signal. %This method has more flexibility in randomizing the selected coefficients over the full signal spectrum. 
%-----------------------------------------------------------------------------------
\begin{figure*}[!t]
%\begin{figure}[!t]
\centering
\includegraphics[width=1.0\textwidth]{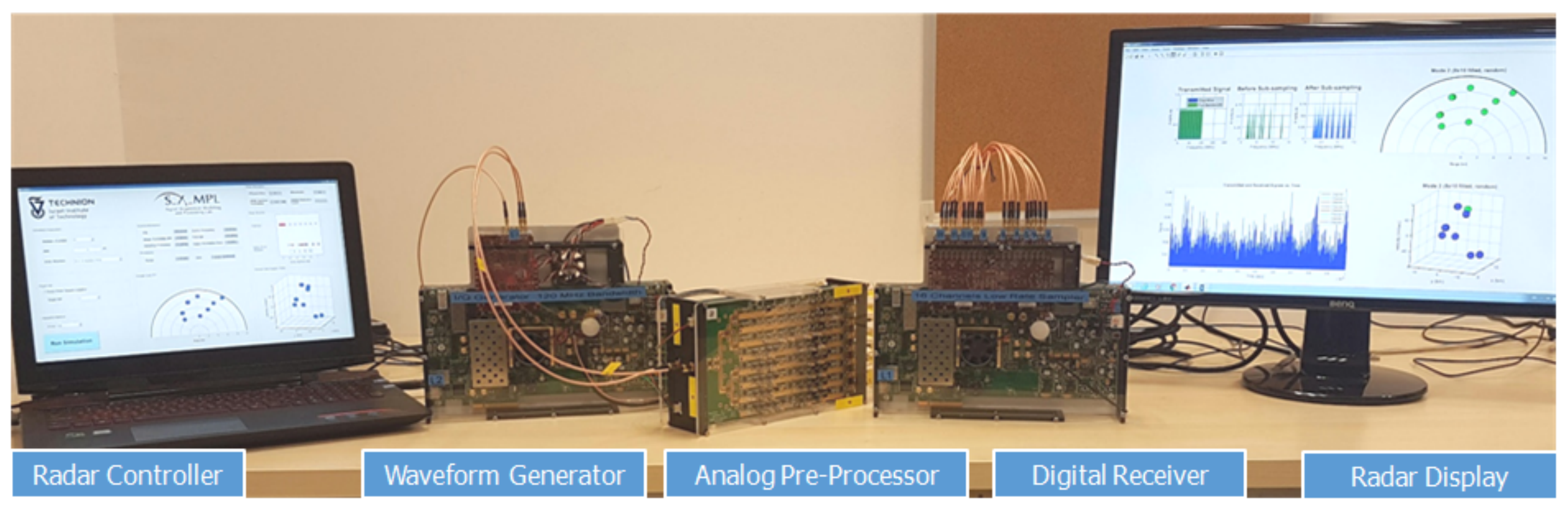}
\caption{CoSUMMeR prototype and its subsystems. The analog pre-processor consists of two cards mounted on the opposite sides of a common chassis.}
\label{fig:mimophoto}
\end{figure*}
%\end{figure}
%-----------------------------------------------------------------------------------

In CoSUMMeR, we employ \textit{foldable multiband sampling} proposed in \cite{mishra2017sub} for a millimeter wave sub-Nyquist receiver. Here, instead of selecting each coefficient randomly from $q$ groups, we choose $N_b$ sets of consecutive coefficients from each group such that the total number of coefficients is still $K$. This implementation needs only eight low-rate ADCs, one per channel. Another advantage of this approach is flexibility of the prototype in selecting the Xampling \textit{slices}. Unlike \cite{baransky2014sub}, the number and spectral locations of slices are not permanently fixed, and they can be changed (within the constraints of aliasing due to subsampling).

Foldable sampling methods suffer from SNR loss due to aliasing of out-band noise. Therefore, the analog front-end must employ filters to reduce undesired leakage of noise. The stop-band attenuation specification of these filters is determined as follows. Let us assume the front-end BPF reduces the out-of-band white noise with power spectral density $N_0$ to $N_{\text{sub}}$ before the translation of the signal to a lower band. For a signal with intermediate frequency (IF) $f_{\text{IF}}$, the subsampling factor is $q = f_{\text{IF}}/f_s$. This subsampling will raise the out-of-band noise power by $2q$ (one spectrum from the negative and another from the positive side) so that the total overlapped noise power density increases to $2qN_{\text{sub}}$. If the overlapped noise power $2qN_{\text{sub}}$ is less than the additive white noise power $N_0$, then the receiver performance remains essentially unaffected. Now, consider a band-limited signal with power spectral density $P_s$. Then, the SNR of the signal sampled at Nyquist rate is given by $\text{SNR}_{\text{Nyq}} = P_s/N_0$. On the other hand, the SNR of the subsampled signal is $\text{SNR}_{\text{sub}} = P_s/(N_0 + 2qN_{\text{sub}})$. The degradation of SNR in dB due to out-band noise is $\text{SNR}_{\text{loss}} = 10\log_{10}\left(\text{SNR}_{\text{Nyq}}/\text{SNR}_{\text{sub}}\right) = 10\log_{10}\left({1 + \frac{2qN_{\text{sub}}}{N_0}}\right)$. As we show in Section~\ref{subsec:app}, analog filters with reasonably high stop-band attenuation are sufficient to mitigate this SNR loss.
%-----------------------------------------------------------------------------------
 \begin{table}[t]
 \centering
 \caption{CoSUMMeR prototype technical specifications}
 \label{tbl:techmodes}
 \vspace{13pt}
 	\begin{tabular}{ l | c | c | c | c}
 		\hline
          \noalign{\vskip 1pt}    
          	Parameters & Mode 1 & Mode 2 & Mode 3 & Mode 4\\[1pt]
 		\hline
 		\hline
         \noalign{\vskip 1pt}    
 		   	\#Tx, \#Rx & 8,10 & 8,10 & 4,5 & 8,10\\[1pt]
 		   	Element placement & Uniform & Random & Random & Random\\[1pt]
 		   	Equivalent aperture & 8x10 & 8x10 & 8x10 & 20x20\\[1pt]            
             Angular resolution (sine of DoA) & 0.025 & 0.025 & 0.025 & 0.005\\[1pt]
 		\hline
         \noalign{\vskip 1pt}    		            
 	        Range resolution & \multicolumn{4}{c}{1.25 m}\\[1pt]
             Signal bandwidth per Tx & \multicolumn{4}{c}{12 MHz (15 MHz including guard-bands)}\\[1pt]
 		   	Pulse width & \multicolumn{4}{c}{4.2 $\mu$s}\\[1pt]
             Carrier frequency & \multicolumn{4}{c}{10 GHz}\\[1pt]
             Unambiguous range & \multicolumn{4}{c}{15 km}\\[1pt]
             Unambiguous DoA & \multicolumn{4}{c}{180$^{\circ}$ (from -90$^{\circ}$ to 90$^{\circ}$)} \\[1pt]   
         \hline
         \noalign{\vskip 1pt}    		            
             PRI & \multicolumn{4}{c}{100 $\mu$s}\\[1pt]
             Pulses per CPI & \multicolumn{4}{c}{10}\\[1pt]
            Unambiguous Doppler & \multicolumn{4}{c}{from $-75$ m/s to $75$ m/s} \\[1pt]
 		\hline
 		%\hline
 	\end{tabular}
 \end{table}
%-----------------------------------------------------------------------------------
 
\subsection{Cognitive Operation and Dynamic Range}
\label{subsec:cog_dr}
Radar receivers typically have dynamic range of tens of dBs to enable detection of targets with a wide range of RCS. In CoSUMMeR, we have an additional constraint of enabling the cognitive operation of the prototype. Therefore, we want to design a prototype with wide dynamic range so that it can handle high in-band power levels. Although there are several techniques to increase the dynamic range of a digital receiver \cite{mishra2012signal,mishra2012frequency}, it is primarily decided by the choice of the ADC. Three ADC design parameters predominantly affect its dynamic range: its Nyquist sampling rate that is determined by the bandwidth of the analog signal to be sampled; number $b$ of digitized bits which should be large to sample with low quantization errors; and the saturation level $P_{\text{sat}}$ dBm under which ADC operation is linear. The lower limit of the dynamic range (in dBm) of the receiver expressed at a given bandwidth BW (typically, 1 MHz) is given by %\citep[p.~163-176]{pace2000},
\begin{align}
\text{DR}_{\text{low}} &= P_{\text{sat}} - 20\log _{10}(2^{b}) - 10\log _{10}\left(\frac{f_s/2}{\text{BW}}\right)\\
 &= P_{\text{sat}} - 6.02b - 10\log _{10}\left(\frac{f_s/2}{\text{BW}}\right).
\end{align}

Unfortunately, numerous factors degrade the ADC's ideal performance resulting in a lower SNR value and higher effective noise figure value \cite{mishra2012frequency}. These errors are often represented by replacing $b$ with the effective number of bits ($E_{\text{NoB}}$) which is much lower than $b$. When quantization noise is also taken into account, the lower limit of the dynamic range (in dBm) is given by %\citep[p.~167]{pace2000},
\begin{align}
\text{DR}_{\text{low}} = P_{\text{sat}} - 1.76 - 6.02E_{\text{NoB}} - 10\log _{10}\left(\frac{f_s/2}{\text{BW}}\right),
\end{align}
so that the dynamic range itself (in dB) is $\text{DR} = P_{\text{sat}} - \text{DR}_{\text{low}}$. Our design goal is to choose an ADC with high $E_{\text{NoB}}$ so that the dynamic range of the receiver does not degrade with subsampling.
%-----------------------------------------------------------------------------------
\begin{figure*}[!t]
\centering
\includegraphics[width=1.0\textwidth]{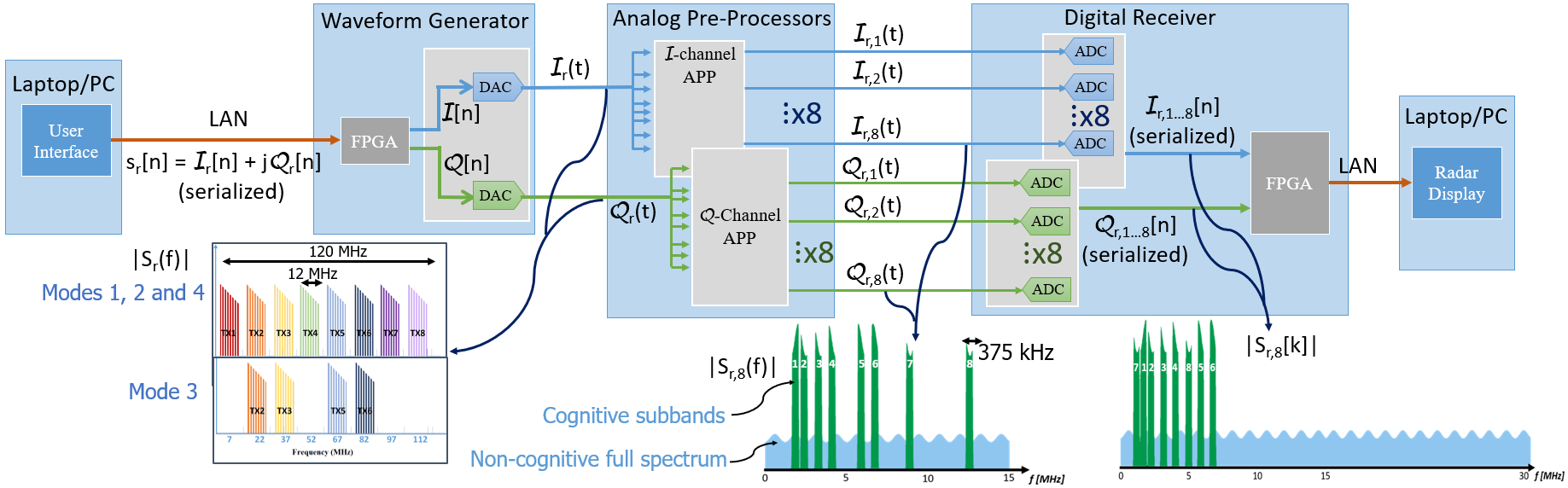}
\caption{Simplified block diagram of the CoSUMMeR prototype. The subscript $r$ represents received signal samples for the $r$th receiver. Wherever applicable, the second subscript corresponds to a transmitter. The square brackets (parentheses) are used for digital (analog) signals. The spectrum $|S_r(f)|$ of the complex received signal for each mode over the $120$ MHz frequency range before separating the echoes corresponding to individual transmitters is illustrated in the bottom left. The locations of cognitive subbands within the non-cognitive full spectrum $|S_{r,8}(f)|$ for a Tx-Rx channel before subsampling is shown in the bottom middle. In the bottom right, the spectrum of the corresponding cognitive signal subsampled at $7.5$ MHz is shown as $|S_{r,8}[k]|$ overlaid on the spectrum of the non-cognitive signal sampled at the Nyquist rate of $30$ MHz.}
\label{fig:mimoblock}
\end{figure*}
%-----------------------------------------------------------------------------------
\section{System Architecture}
\label{sec:sys_desc}
Table \ref{tbl:techmodes} summarizes the technical parameters of our CoSUMMeR prototype for all four array configurations. The desired range and angular resolutions as well as maximum unambiguous range and Doppler velocity requirements are based on some of the common MIMO radars mentioned in Section~\ref{sec:intro}. Based on these specifications, we chose $10$ pulses per CPI with a PRI of $100$ $\mu$s. Each transmit signal has an approximately flat spectrum, over the extent of 12 MHz (one-sided band). The waveforms are separated from each other by a 3 MHz guard-band so that the total bandwidth occupied by all $8$ transmitters is $120$ MHz.

Figure~\ref{fig:mimophoto} shows the CoSUMMeR prototype. It consists of a radar controller, waveform generator, analog pre-processor (APP), digital receiver and data-processor-combined-with-display. Figure~\ref{fig:mimoblock} illustrates the simplified block diagram of the prototype with the signal spectrum at every major stage. We now provide details on each one of the subsystems.
 
\subsection{Radar Controller}
\label{subsec:radar_cont}
%-----------------------------------------------------------------------------------
\begin{figure*}[!t]
%\begin{figure}
\centering
\includegraphics[width=1.0\textwidth]{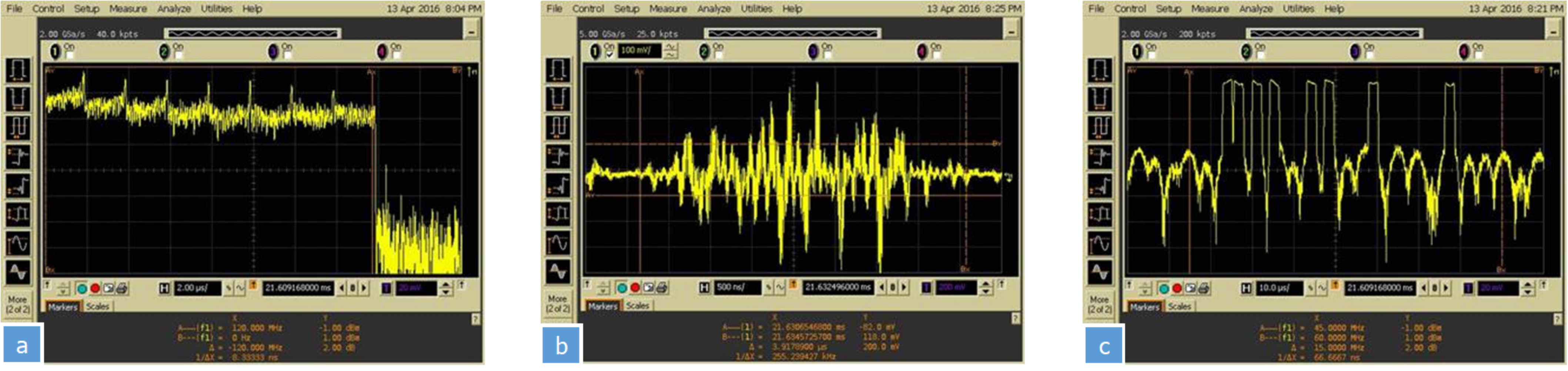}
\caption{Output of the waveform generator for one receive chain. (a) Spectrum over 120 MHz showing echoes from all 8 transmitters. (b) Single transmit waveform in time-domain. (c) Spectrum of single transmit signal showing that only eight 375 kHz slices out of available $12$ MHz bandwidth are transmitted.}
\label{fig:wfgenOutput}
\end{figure*}
%\end{figure}
%-----------------------------------------------------------------------------------
The prototype is configured through the user interface of the radar controller software deployed on a high-end portable server. The software consists of two major components: waveform control and APP control. These components interact with each other and provide status to the data processor. The waveform control allows the user to select the target scenario, array constellations (or modes), and SNR levels of the received baseband signals. The option to operate Mode 3 cognitively is also provided, along with the facility to auto-calibrate the prototype using Built-In-Test-Equipment (BITE) and Built-In-System-Test (BIST) signals \cite{mishra2012frequency}. The APP control allows the controller to skip individual Tx channels while operating in spatial sub-Nyquist mode. All hardware devices are individually powered and the controller communicates with them over an Ethernet link. 

\subsection{Waveform Generator}
\label{subsec:wfgen}
The user selects the prototype mode from the control interface and passes the control triggers to the waveform generator card. The waveform generator is off-the-shelf Xilinx VC707 evaluation board that is custom fit with a 4DSP FMC204 16-bit DAC mezzanine card. The waveforms are stored as digital $\mathcal{I}/\mathcal{Q}$ pairs at baseband with a sample rate of $250$ MHz for each channel. The transmit waveform is downloaded to the waveform generator's onboard $1$ GB DDR3 memory via Ethernet interface either through a server or single-board computer. The FPGA device then reads out the pre-stored waveform from the memory and employs $8$ Gbps Serializer/Deserializer (SerDes) device to transfer it to two separate $16$-bit DACs, one each for $\mathcal{I}$ and $\mathcal{Q}$ samples. The DACs then interpolate and convert the stored waveform to an analog baseband signal at a sample rate of $1$ Gsps. This process continues until all the $\mathcal{I}$-$\mathcal{Q}$ pairs of received waveforms corresponding to each Tx-Rx channels are fetched from memory. Each of the $\mathcal{I}$ and $\mathcal{Q}$ analog signals are then passed on to their respective analog pre-processor cards. The waveform generator also produces timing signals such as system clock and triggers to indicate the beginning of each PRI. The latter is used by the digital receiver to begin the sampling operation when a new range-time profile is received.

Figure~\ref{fig:wfgenOutput}a shows the output of the waveform generator for one receiver as seen on a spectrum analyzer. The echoes corresponding to all eight transmitters are present. Together, this signal is spread over $120$ MHz. The time and frequency domain representations of a signal corresponding to a single Tx-Rx channel are provided in Figs~\ref{fig:wfgenOutput}b-c. Since the transmission is cognitive, the signal is restricted to only $N_b = 8$ narrow subbands, each of $375$ kHz bandwidth. These cosets are chosen after extensive software simulations such that they provide low mutual coherence.

\subsection{APPs}
\label{subsec:app}
%-----------------------------------------------------------------------------------
\begin{figure}
\centering
\includegraphics[width=1.0\columnwidth]{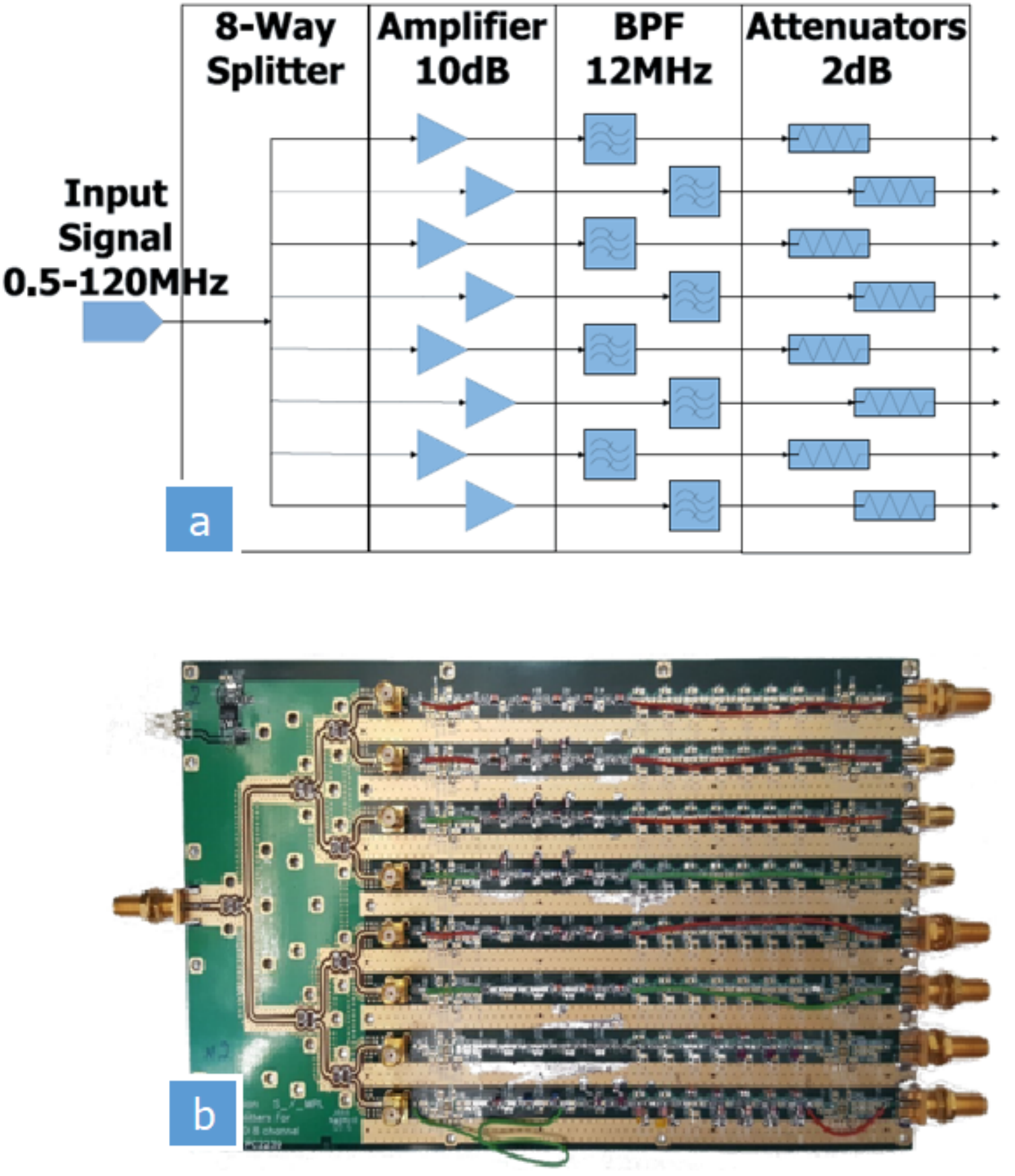}
\caption{(a) Block diagram of the APP chain. (b) An APP card showing the baseband lumped circuit filters.}
\label{fig:appBlockPhoto}
\end{figure}
%-----------------------------------------------------------------------------------
%-----------------------------------------------------------------------------------
\begin{table*}[t]
\centering
\renewcommand{\arraystretch}{1.2}
\caption{Specifications of APP filters}
\label{tbl:appfilters}
	\begin{tabular}{ l | c | c | c | c | c | c | c}
		\hline
         \noalign{\vskip 1pt}    
         	Channel & Center Frequency & Passband & Type & Order & Passband Ripple (dB) & Stopband Attenuation & Equalizer\\[1pt]
		\hline
		\hline
        \noalign{\vskip 1pt}    
		   	1 & 7 MHz & 0-13.4 MHz & Low-pass elliptic & 7 & 0.01 & 30dB$@$16MHz & No\\
            \hline
		   	2 & 22 MHz & 16-28 MHz & Band-pass elliptic & 5 & 0.4 & \begin{tabular}[c]{@{}c@{}}30 dB $@$ 16 MHz \\30 dB $@$ 31MHz \end{tabular} & Yes\\
            \hline
   		   	3 & 37 MHz & 31-43 MHz & Band-pass elliptic & 5 & 0.4 & \begin{tabular}[c]{@{}c@{}}30 dB $@$ 28 MHz \\30 dB $@$ 46 MHz \end{tabular} & Yes\\
            \hline
		   	4 & 52 MHz & 46.5-59.5 MHz & Low-pass \& high-pass elliptic & \begin{tabular}[c]{@{}c@{}}9 (low-pass), \\9 (high-pass) \end{tabular} & 0.1 & \begin{tabular}[c]{@{}c@{}}30 dB $@$ 43 MHz \\30 dB $@$ 61 MHz \end{tabular} & No\\
            \hline
			5 & 67 MHz & 61-73 MHz & Low-pass \& high-pass elliptic & \begin{tabular}[c]{@{}c@{}}9 (low-pass), \\9 (high-pass) \end{tabular} & 0.1 & \begin{tabular}[c]{@{}c@{}}30 dB $@$ 58 MHz \\30 dB $@$ 76 MHz \end{tabular} & No\\
            \hline
            6 & 82 MHz & 76-88 MHz & Low-pass \& high-pass elliptic & \begin{tabular}[c]{@{}c@{}}9 (low-pass), \\9 (high-pass) \end{tabular} & 0.1 & \begin{tabular}[c]{@{}c@{}}30 dB $@$ 73 MHz \\30 dB $@$ 91 MHz \end{tabular} & No\\
            \hline
            7 & 97 MHz & 91-103 MHz & Low-pass \& high-pass elliptic & \begin{tabular}[c]{@{}c@{}}9 (low-pass), \\9 (high-pass) \end{tabular} & 0.1 & \begin{tabular}[c]{@{}c@{}}30 dB $@$ 88 MHz \\30 dB $@$ 106 MHz \end{tabular} & No\\
            \hline
            8 & 112 MHz & 106-118 MHz & Low-pass \& high-pass elliptic & \begin{tabular}[c]{@{}c@{}}9 (low-pass), \\11 (high-pass) \end{tabular} & 0.1 & \begin{tabular}[c]{@{}c@{}}30 dB $@$ 103 MHz \\30 dB $@$ 121 MHz \end{tabular} & No\\
		\hline
		\hline
	\end{tabular}
\end{table*}
%-----------------------------------------------------------------------------------
\begin{figure*}[!t]
\centering
\includegraphics[width=0.8\textwidth]{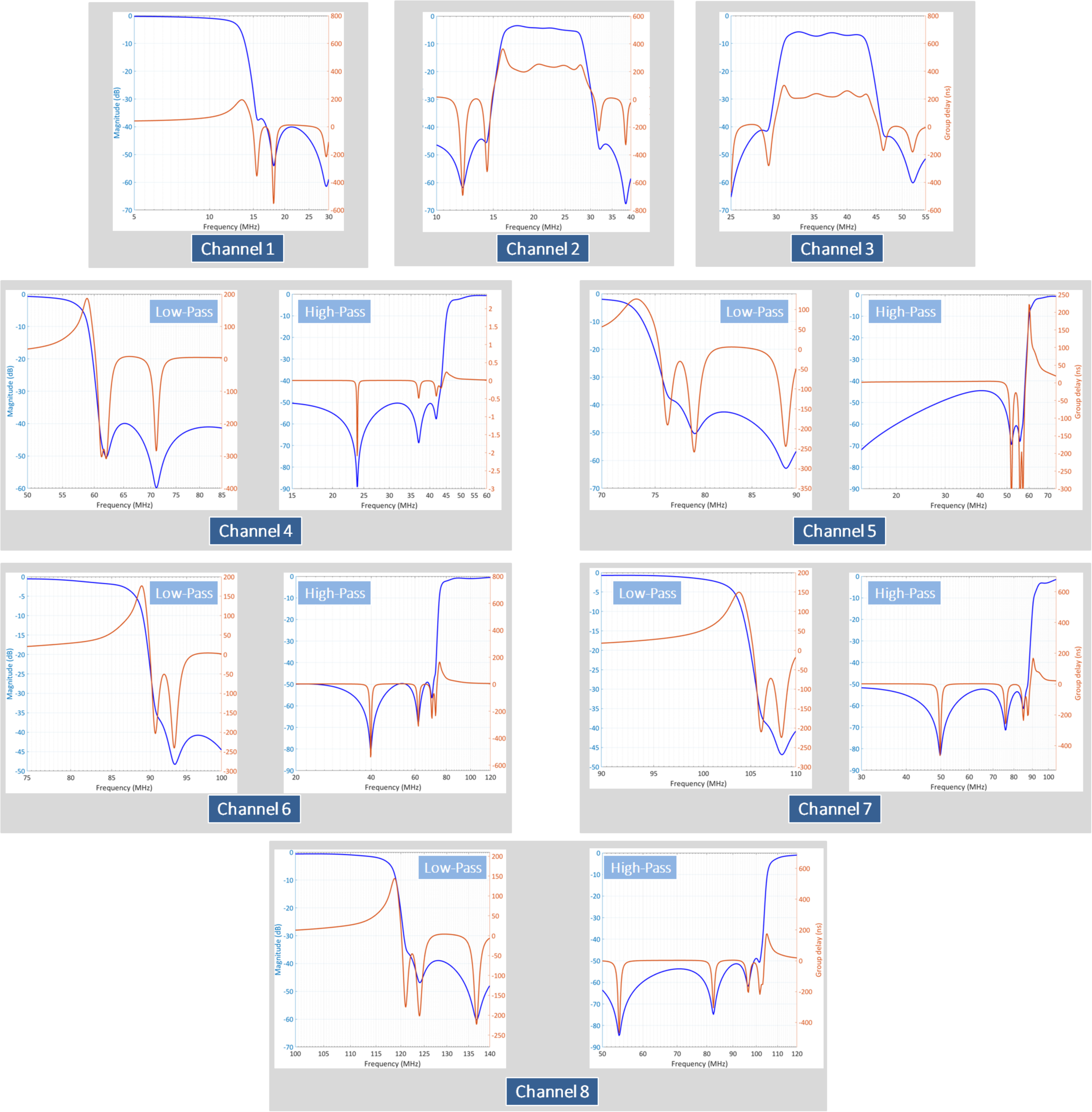}
\caption{Magnitude and group delay responses of analog filters. Filters for channels 4-8 are realized separately as low- and high-pass filters in the hardware.}
\label{fig:app_responses}
\end{figure*}
%-----------------------------------------------------------------------------------
A custom-built APP (Fig.~\ref{fig:appBlockPhoto}) splits the 120 MHz baseband analog signal from the waveform generator into 8 channels. The 9 dB attenuation due to an 8-channel Wilkinson splitter is compensated with the use of a 10 dB amplifier for each channel. The signal corresponding to each transmitter is then filtered using BPFs each with 12 MHz passband. We use two separate APP cards, for $\mathcal{I}$ and $\mathcal{Q}$ channels, mounted on the opposite sides of a single chassis (Fig.~\ref{fig:mimophoto}). 

We designed the analog filters for each Tx-Rx channel to obtain nearly $30$ dB stop-band attenuation. As we explain below, our digital receiver subsamples the analog signal at $7.5$ MHz while the Nyquist rate is $30$ MHz. This yields a subsampling factor of $q=4$. With these specifications, the stop-band attenuation of $30$ dB (or $N_0/N_{\text{sub}} = 1000$, on linear scale) leads to an $\text{SNR}_{\text{loss}}$ of only $0.035$ dB, hugely mitigating the effect of out-band noise. All analog filters are elliptic because, for any given order, the elliptic filter gives a much higher rate of attenuation in the transition band. In our case, the guard-band being $3$ MHz, the filter response must reach stop-band attenuation of $30$ dB at most $3$ MHz away from the passband cut-off frequency. At the same time, the phase distortion or group delay in elliptic filters is the worst among standard responses such as Butterworth, Chebyshev, and Bessel filters \cite{valkenburg1982analog}. We correct this distortion digitally in the receiver through extensive calibration of the entire analog chain.

Table~\ref{tbl:appfilters} lists the specifications of all analog filters employed in the CoSUMMeR prototype. Only the first transmit channel uses a low-pass elliptic filter because it is difficult to practically realize a bandpass filter with a passband close to zero. The second and third channel filters are followed by tunable equalizers to correct the imbalance in gain. All other BPFs are realized by a combination of low- and high-pass elliptic filters with an overall passband ripple of 0.1 dB. Figure~\ref{fig:app_responses} shows the theoretical magnitude and group delay response of each filter. Most of the responses achieve 30 dB attenuation in a transition band less than $3$ MHz. Finally, as shown in Fig.~\ref{fig:appBlockPhoto}a, $2$ dB attenuators are placed after each BPF in order to avoid saturating the digital receiver that follows APPs. 

\begin{figure}
\centering
\includegraphics[width=1.0\columnwidth]{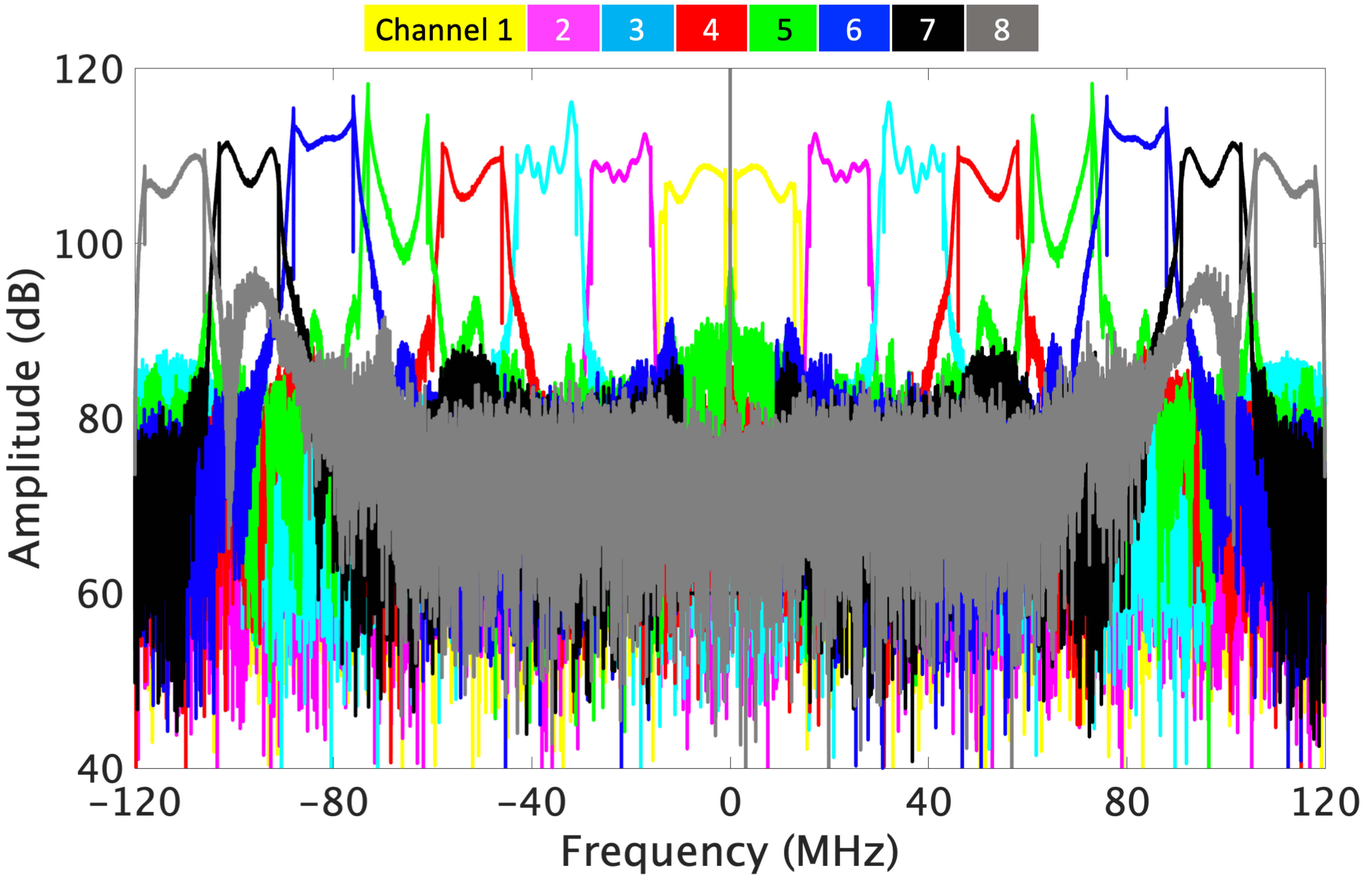}
\caption{Two-sided sampled magnitude response of all eight channels when the input is a BITE signal with an approximately flat spectrum across $120$ MHz range.}
\label{fig:bite_response}
\end{figure}

Figure~\ref{fig:bite_response} shows the response of all eight filters as sampled by the digital receiver. The injected signal at the waveform generator was a standard BITE waveform (with flat spectrum over $120$ MHz). We note that our design of analog filters is quite robust and provides excellent isolation of adjacent channels.
%-----------------------------------------------------------------------------------
%-----------------------------------------------------------------------------------
\subsection{Digital Receivers}
\label{subsec:drx}
\begin{figure}[!t]
\centering
\subfloat[Before subsampling]{%
  \includegraphics[width=4.1cm]{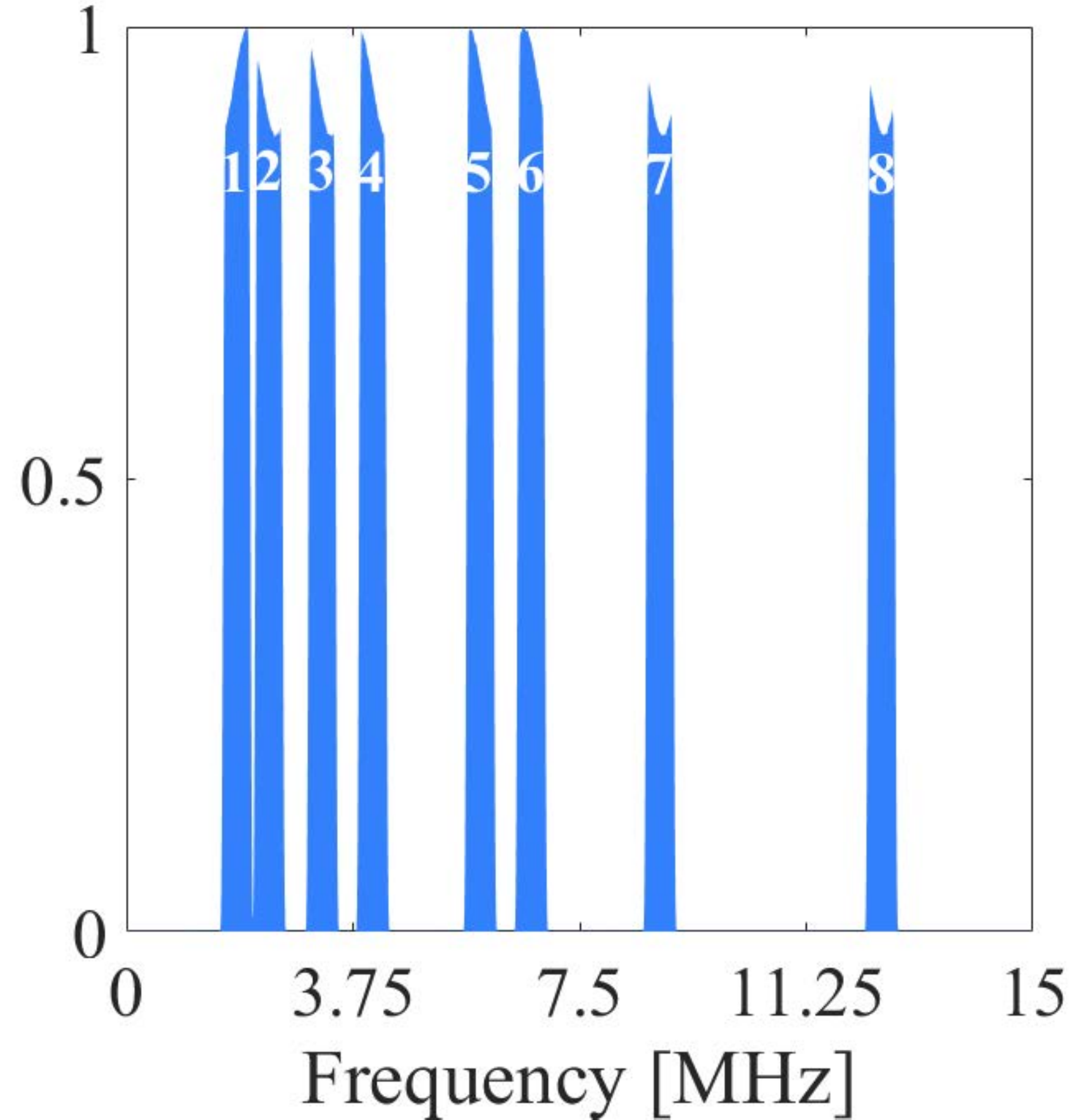}%
  \label{fig:beforesub}%
}\qquad
\subfloat[After subsampling]{%
  \includegraphics[width=4.1cm]{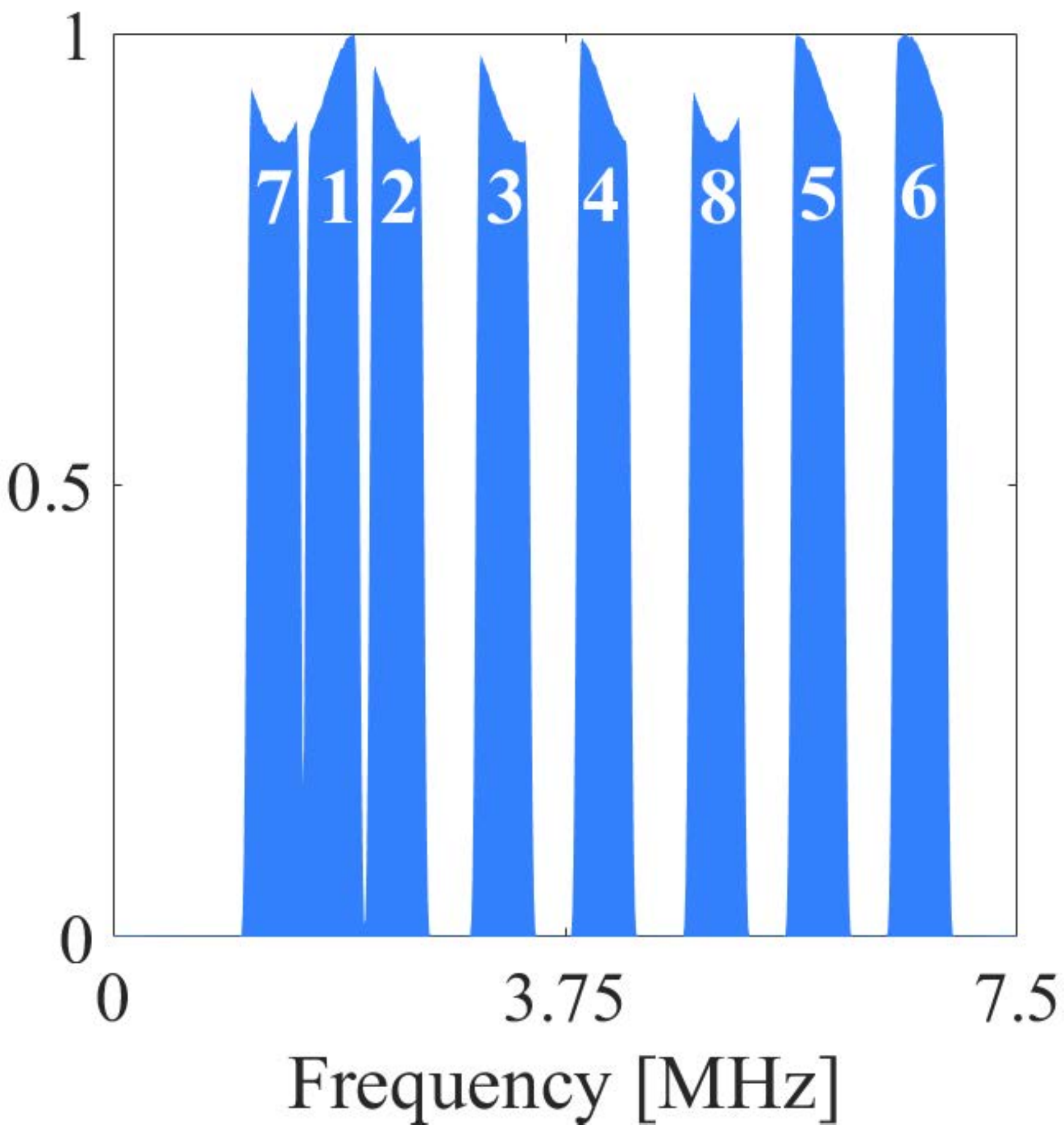}%
  \label{fig:aftersub}%
}
\caption{The normalized one-sided spectrum of one channel of a given receiver (a) before and (b) after subsampling with a 7.5 MHz ADC. Each of the subbands span 375 kHz and is marked with a numeric label. In a non-cognitive processing, the signal occupies the entire 15 MHz spectrum before sampling.}
\label{fig:slicespec}
\end{figure}

The digital receiver consists of a single Xilinx VC707 evaluation board with two eight-channel 4DSP FMC168 digitizer daughter cards, one each for $\mathcal{I}$ and $\mathcal{Q}$ signals. The channelized $\mathcal{I}/\mathcal{Q}$ analog signals are then digitized using low-rate 16-bit ADCs in a digital receiver card. As shown in Fig.~\ref{fig:beforesub}, the cognitive radar signal occupies only certain subbands in a 15 MHz band of a single transmitter. Here, the sliced transmit signal has eight subbands each of width 375 kHz with frequency ranges 1.63-2, 2.16-2.53, 3.05-3.42, 3.88-4.25, 5.66-6.03, 6.51-6.88, 8.64-9.01 and 12.32-12.69 MHz before subsampling. The resulting coherence \cite{eldar2012compressed} for this selection of Fourier coefficients is 0.42. The total signal bandwidth is $0.375 \times 8 = 3$ MHz. This signal is subsampled at 7.5 MHz and, as shown in Fig.~\ref{fig:aftersub}, there is no aliasing between different subbands. 

The 4DSP FMC168 employs Texas Instruments ADS42LB69 ADC that has $E_{\text{NoB}} = 11.85$ and $P_{\text{sat}}=\pm 10$ dBm. This results in a dynamic range (in dB) of,
\begin{align}
\text{DR} = -1.76 + 6.02(11.85) + 10\log _{10}\left(\frac{7.5/2}{1}\right) = 75.32,
\end{align}
with the lower limit of $\text{DR}_{\text{low}} = -10-75.32 = -85.32$ dBm which is close to the digital noise floor of the receiver. This range is sufficient for us to experiment with various power levels during the cognitive mode. By default, we distribute power equally in all the subbands which are of equal bandwidth, although other considerations such as presence of interference may also guide power allocation in practice \cite{mishra2017performance}.

The sampled data is transferred to an FPGA using $1.2$ Gbps SerDes. The FPGA then writes the data to a digital first-in-first-out (FIFO) buffer from where an RJ-45 controller reads and transfers the signals to a data processor over the Ethernet link. 
%-----------------------------------------------------------------------------------
%-----------------------------------------------------------------------------------
\subsection{Data Processors and Radar Display}
%-----------------------------------------------------------------------------------
\begin{table*}[t]
\centering
\renewcommand{\arraystretch}{1.2}
\caption{CoSUMMeR Prototype: Comparison of Resource Reduction}
\label{tbl:cogsum_red}
	\begin{tabular}{p{4.5cm}|p{1.6cm}p{2.1cm}p{0.85cm}||p{2.2cm}p{2.1cm}p{0.85cm}}
	\hline\noalign{\smallskip}
	Resource & Nyquist Mode 1 & Sub-Nyquist Mode 3 & Reduction & Nyquist $20\times20$ array & Sub-Nyquist Mode 4 & Reduction\\
	\noalign{\smallskip}
	\hline
	\noalign{\smallskip}
	Bandwidth per Tx (including guard-bands) & $15$ MHz  & $3$ MHz & $80$\% & $15$ MHz & $3$ MHz & $80$\% \\
	Bandwidth per Tx (excluding guard-bands) & $12$ MHz  & $3$ MHz & $75$\% & $12$ MHz  & $3$ MHz & $75$\% \\
	Temporal sampling rate per channel & $30$ MHz & $7.5$ MHz & $75$\% & $30$ MHz & $7.5$ MHz & $75$\%\\
	Spatial sampling rate & $8\times 10$  & $4\times 5$ & $50$\% & $20\times 20$  & $8\times 10$ & $55$\% \\
	Tx/Rx hardware channels & $80$  & $20$ & $75$\% & $400$ & $80$ & $80$\% \\
    Total Tx bandwidth (including guard-bands) & $120$ MHz & $12$ MHz & $90$\% & $300$ MHz & $24$ MHz & $92$\% \\
    Total Tx bandwidth (excluding guard-bands) & $96$ MHz & $12$ MHz & $87.5$\% & $240$ MHz & $24$ MHz & $90$\% \\
	\noalign{\smallskip}\hline\noalign{\smallskip}
	\end{tabular}
\end{table*}
%-----------------------------------------------------------------------------------
The data processor is a 64-bit Desktop server that receives the sampled data from the receiver and performs signal reconstruction by implementing the algorithm in \cite{cohen2018summer} in real-time. The number of Fourier coefficients corresponding to each transmitter bandwidth of $15$ (or $12$) MHz is 1500 while that for a single subband of $375$ kHz bandwidth is $38$. Therefore, the data processor uses a total of $8\times38=304$ Fourier coefficients during the reconstruction process.

The detection results of sub-Nyquist signal processing are shown on a radar display (see Fig.~\ref{fig:mimophoto}) in polar range-azimuth and Cartesian range-angle-Doppler plots. Here, individual channels can also be examined through their time-domain or A-scope plots. The display also stores the previous three results to allow comparison with the current configuration.

\subsection{Resource Reduction}
\label{subsec:res_red}
In Fig.~\ref{fig:mimoblock}, we overlay the spectra of Fig.~\ref{fig:slicespec} over the full bandwidth of the non-cognitive signal before subsampling. A non-cognitive signal occupies the entire 15 MHz spectrum requiring a Nyquist sampling rate of 30 MHz. On the other hand, our digital receiver samples at $7.5$ MHz. Therefore, use of cognitive transmission enables temporal sampling reduction by a factor of $4$ ($=30$ MHz$/7.5$ MHz) for each channel. Depending on whether the guard-bands of the non-cognitive transmission are included in the computation or not, the effective signal bandwidth for each channel is reduced by a factor of $5$ ($=15$ MHz$/3$ MHz) or $4$ ($=12$ MHz$/3$ MHz), respectively. The sub-Nyquist array in Mode 3 employs only half the antenna elements than the Nyquist array in Mode 1. In other words, the spatial sampling rate in Mode 3 is reduced by $50\%$ when compared with Mode 1 or 2. If we account for both spatial and spectral sampling reduction in Mode 3, then we use a total one-eighth of the Nyquist sampling rate and one-tenth of the Nyquist signal bandwidth (guard-bands included). The sampling rate reduction is, therefore, seven-eighth or $87.5\%$ in Mode 3. In terms of the hardware cost, the receiver processes $80$ and $20$ Tx-Rx channels in $8\times10$ and $4\times5$ arrays, respectively. Thus, the hardware resources are reduced by $75\%$ in Mode 3.

Similar comparisons of Mode 4 can be made with its Nyquist equivalent $20\times20$ array. Here, the reduction in temporal sampling rate and per transmitter bandwidth is the same as before. Since Mode 4 employs a total of $18$ antenna elements compared to $40$ of the virtual array, the sampling rate reduction is $55\%$. The corresponding savings in hardware due to reduction of Tx-Rx channels from $400$ in the virtual array to $80$ in Mode 4 is substantial ($80\%$). The spatial sub-Nyquist rate reduction leads to a significantly low usage of transmit bandwidth in Mode 4. Table~\ref{tbl:cogsum_red} summarizes these comparisons.

In terms of system complexity, our design for $80$ Tx-Rx channels is accomplished using $8$ ADCs, $8$ BPFs, and $8$ analog channels. A non-cognitive Nyquist design would require $80$ units of each of these components. For a non-cognitive sub-Nyquist design that serializes all receive data-streams and employs subsampling, $88$ BPFs - one for separating each of the Tx signals and another $8$ to separate each of the subbands in $10$ receivers - and $8$ ADCs are required. If subsampling is forfeited, then serialization of receive data in a non-cognitive sub-Nyquist design would consume $64$ ADCs (one for each of the $8$ subbands in $8$ Tx) and $72$ BPFs ($64$ each of the ADCs and $8$ more to separate each Tx signal).

\section{Experimental Results}
\label{sec:results}
We evaluated the performance of CoSUMMeR through hardware experiments. The waveform generator produced $P=10$ pulses at a PRI of $100$ $\mu$s and we compared the signal reconstruction results of all modes for identical target scenarios. Cognitive transmission was operated at the near saturation level of the devices. While configuring the prototype for the non-cognitive modes, we decreased the signal strength accordingly. 

In the following, we present radar display outputs of three experiments. Here, a successful detection (circle with dark boundary and no fill) occurs when the estimated target is within one range cell, one azimuth bin and one Doppler bin of the ground truth (circle with light fill and no boundary); otherwise, the estimated target is labeled as a false alarm (circle with dark fill). Only for the purposes of clear illustration, we have magnified the circles; the exact location of the targets should be taken as the centre of the circles.

\begin{figure}[!t]
\centering
\includegraphics[width=1.0\columnwidth]{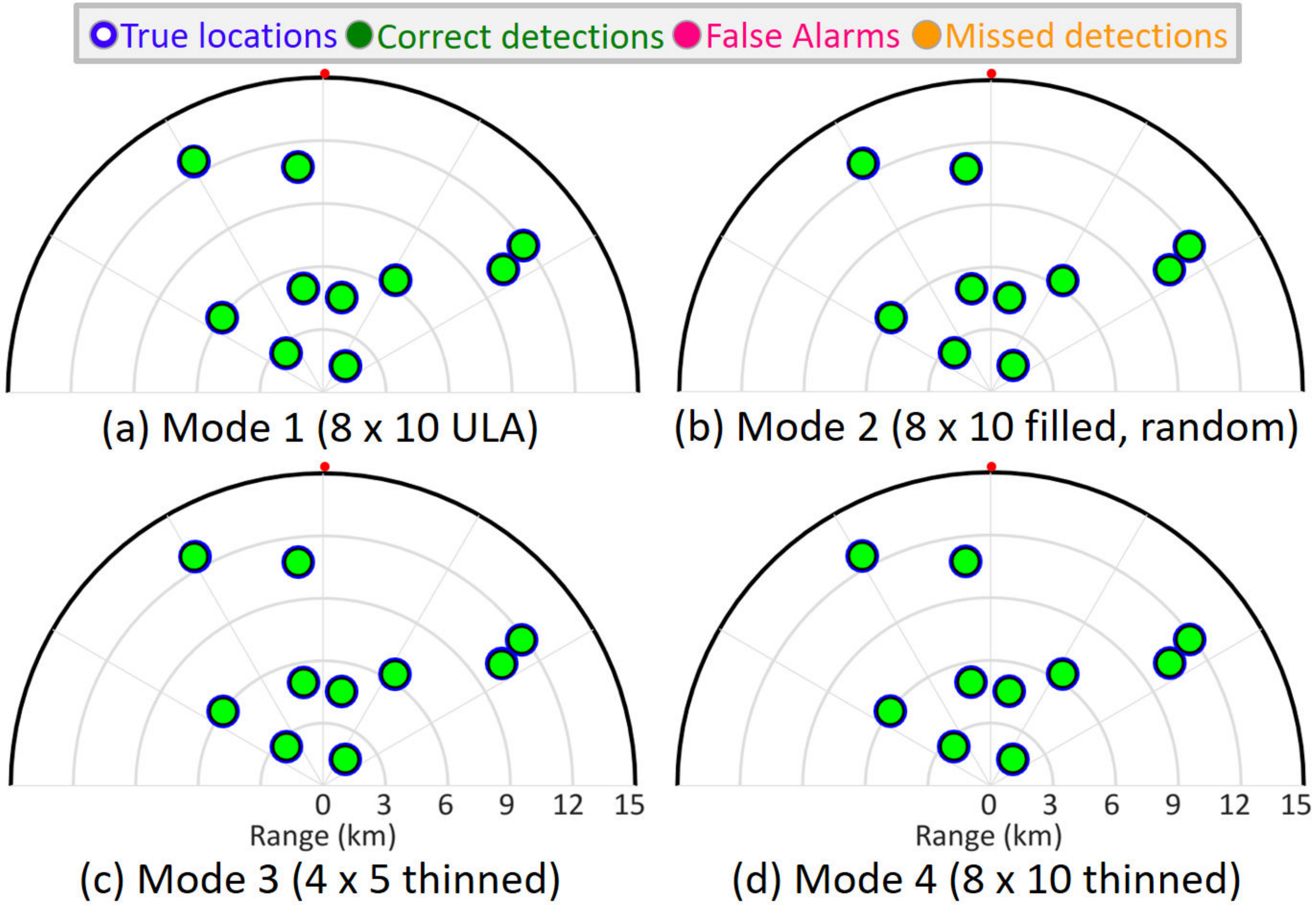}
\caption{Plan Position Indicator (PPI) display of results for Modes 1 through 4. The origin is the location of the radar. The dark dot indicates the north direction relative to the radar. Positive (negative) distances along the horizontal axis correspond to the east (west) of the radar. Similarly, positive (negative) distances along the vertical axis correspond to the north (south) of the radar. The estimated targets are plotted over the ground truth.}
\label{fig:sameperf}
\end{figure}

\begin{figure}[!t]
\centering
\includegraphics[width=1.0\columnwidth]{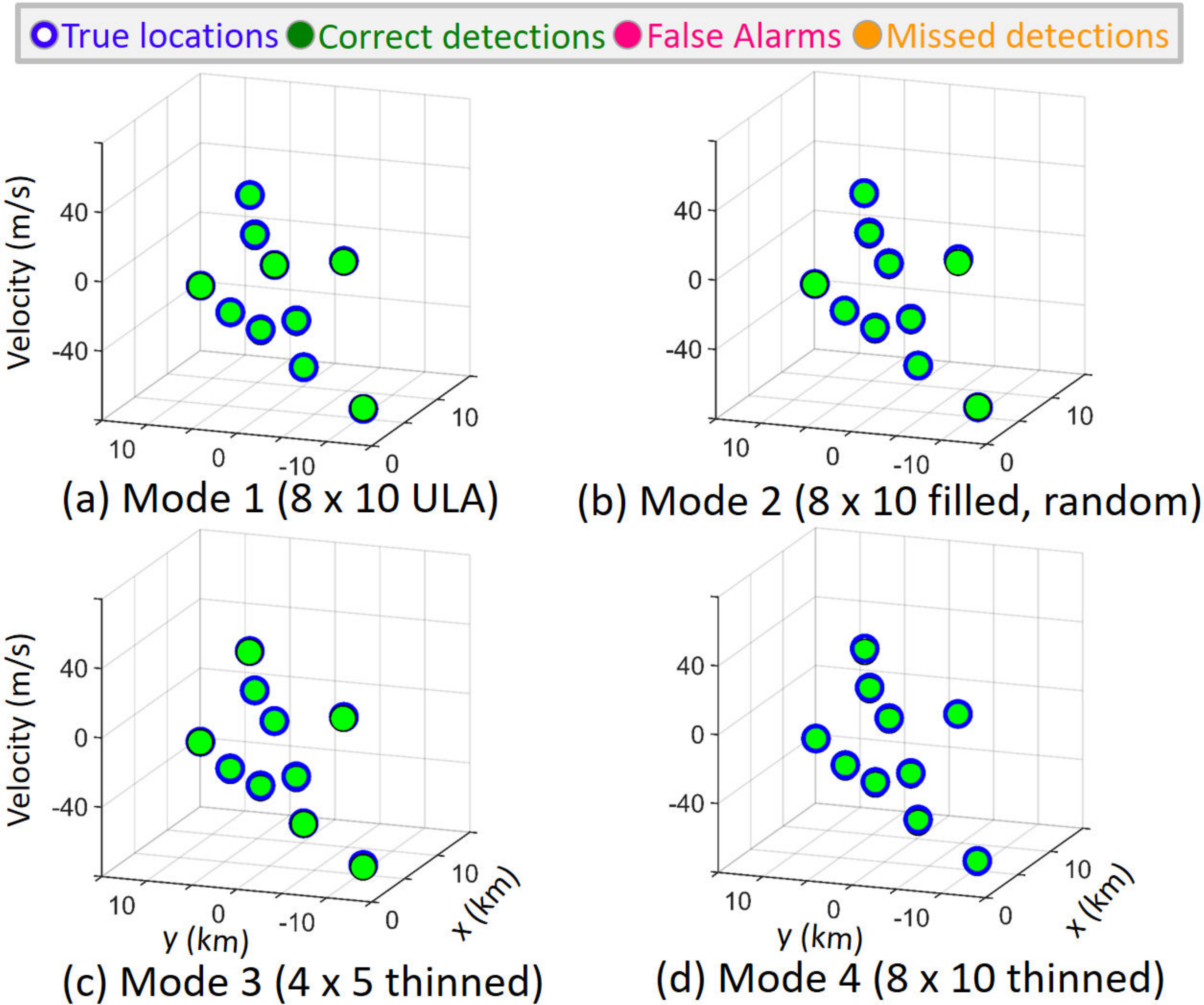}
\caption{Range-Azimuth-Doppler map for the target configurations shown in Fig.~\ref{fig:sameperf}. The lower axes represent Cartesian coordinates of the polar representation of the PPI plots from Fig.~\ref{fig:sameperf}. The vertical axis represents the Doppler spectrum.}
\label{fig:sameperf_ddm}
\end{figure}

\begin{figure}[!t]
\centering
\includegraphics[width=1.0\columnwidth]{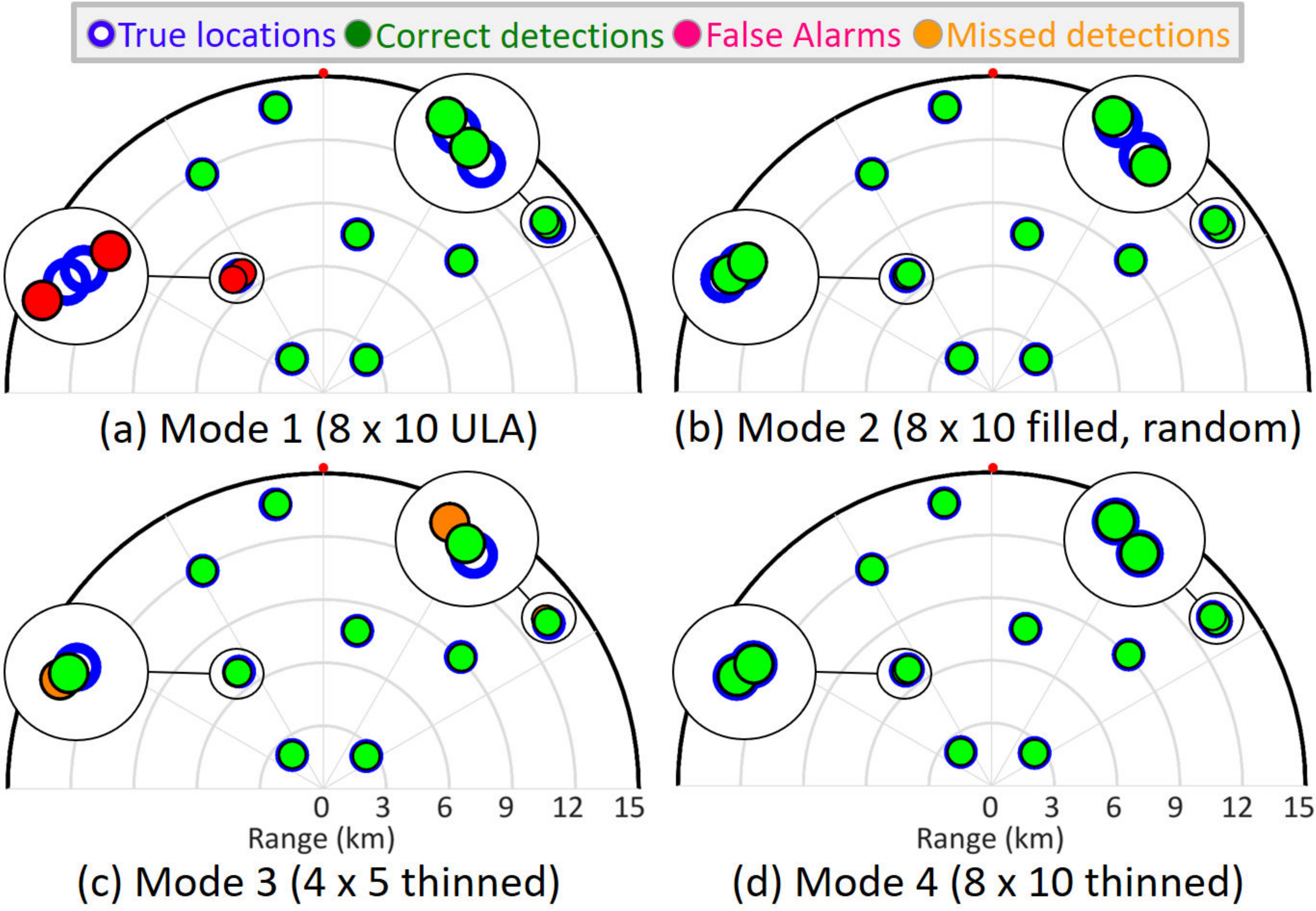}
\caption{PPI plots as in Fig.~\ref{fig:sameperf}, but for a closely-spaced target scenario. The inset plots show the selected region in each PPI display on a magnified scale.}
\label{fig:diffperf}
\end{figure}

\begin{figure}[!t]
\centering
\includegraphics[width=1.0\columnwidth]{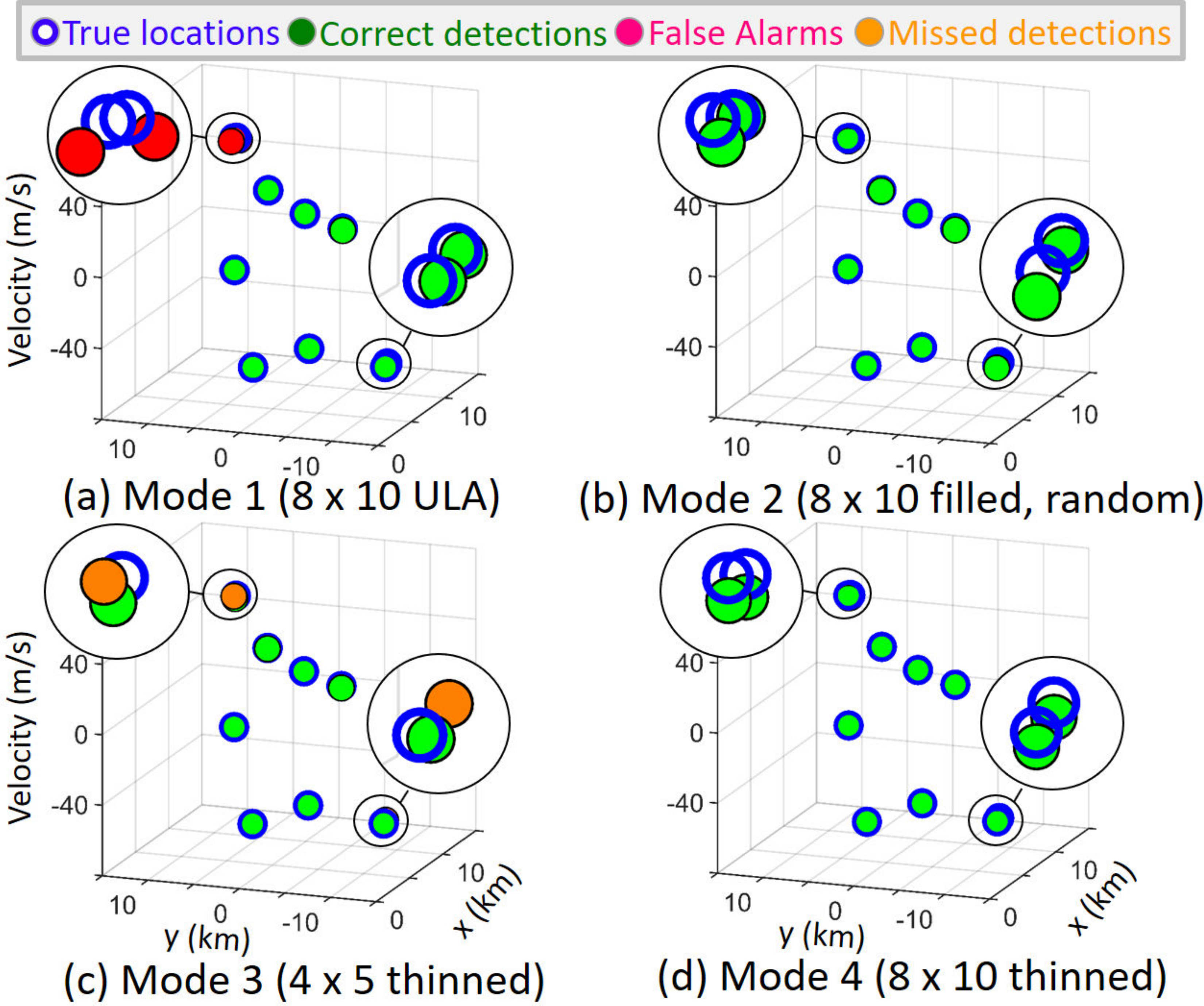}
\caption{Range-Azimuth-Doppler maps as in Fig.~\ref{fig:sameperf_ddm}, but for a closely-spaced target scenario. The inset plots show the selected region in each map on a magnified scale.}
\label{fig:diffperf_ddm}
\end{figure}

\begin{figure}[!t]
\centering
\includegraphics[width=1.0\columnwidth]{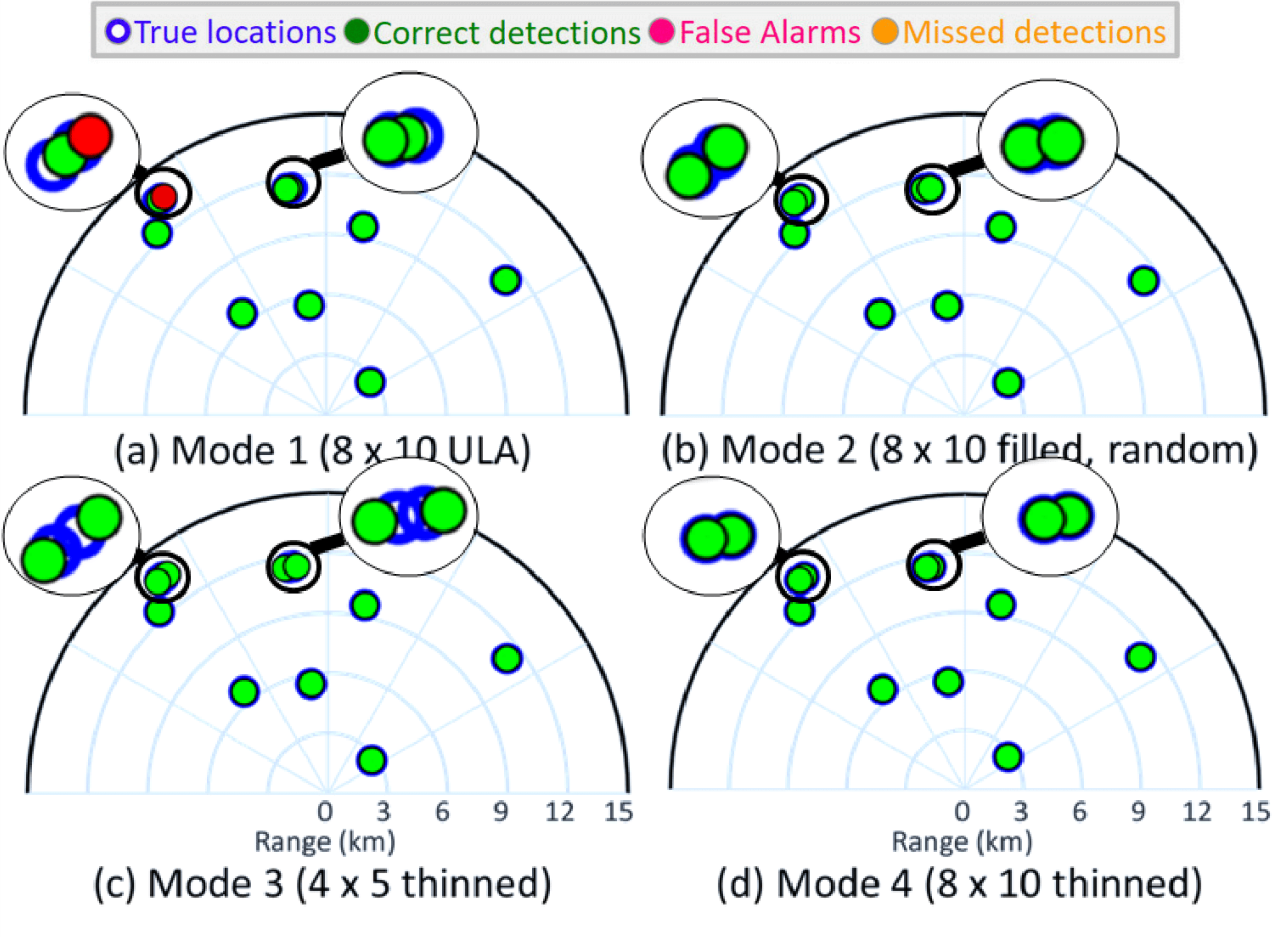}
\caption{PPI plots as in Fig.~\ref{fig:sameperf}, but only Mode 3 is operating cognitively. All modes have the same overall transmit power per transmitter. The inset plots show the selected region in each PPI display on a magnified scale.}
\label{fig:cogperf}
\end{figure}

\begin{figure}[!t]
\centering
\includegraphics[width=1.0\columnwidth]{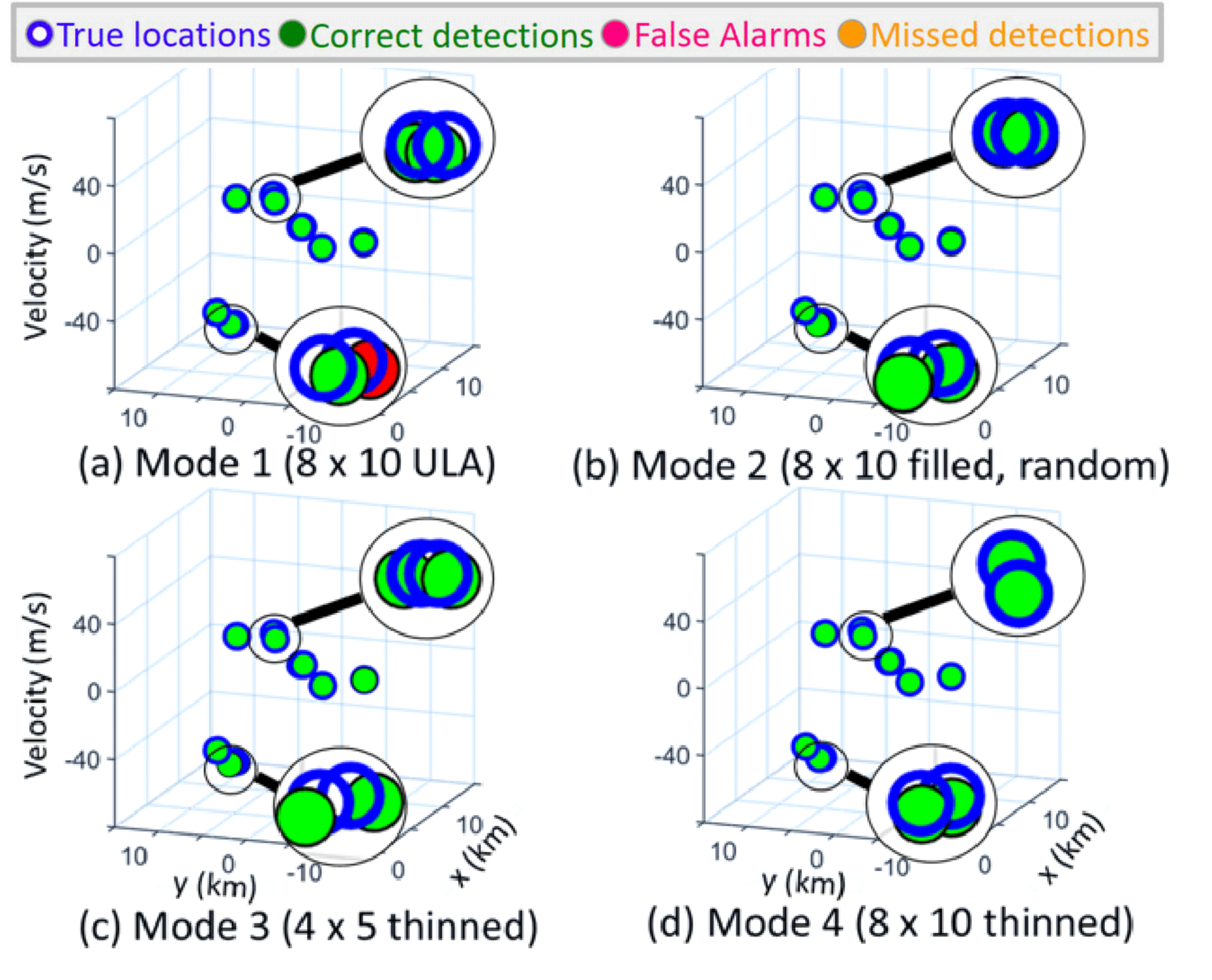}
\caption{Range-Azimuth-Doppler maps as in Fig.~\ref{fig:sameperf_ddm}, but only Mode 3 is operating cognitively. All modes have the same overall transmit power per transmitter. The inset plots show the selected region in each map on a magnified scale.}
\label{fig:cogperf_ddm}
\end{figure}

\begin{figure}[!t]
\centering
\includegraphics[width=0.8\columnwidth]{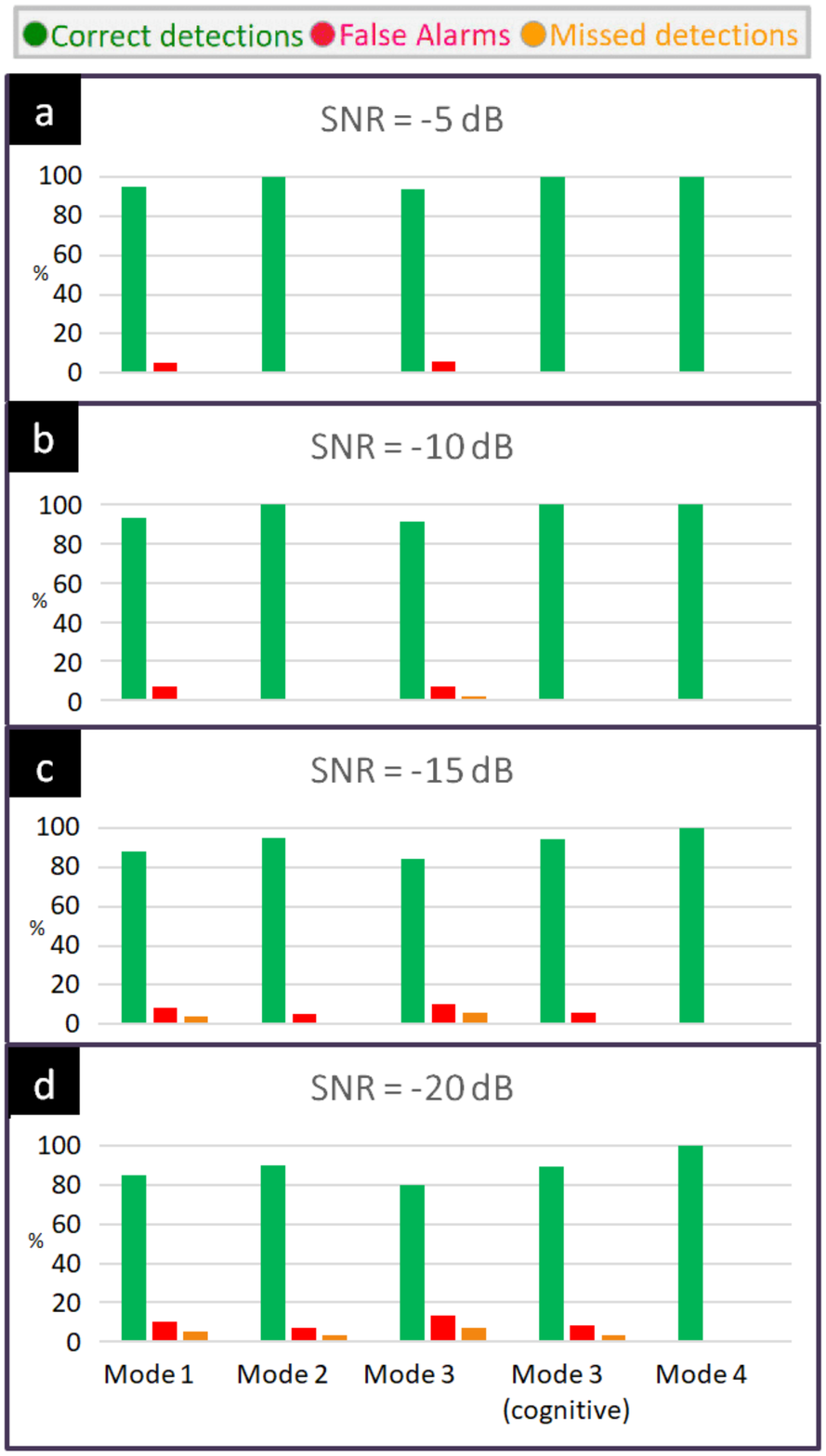}
\caption{Statistical performance of different array configurations in a CoSUMMeR prototype over $100$ realizations of randomly spaced target scenes.}
\label{fig:perfStats}
\end{figure}

\noindent\textbf{Randomly spaced targets}: In the first experiment, when the angular spacing (in terms of the sine of azimuth) between any two targets was greater than $0.025$ and the signal SNR = $-8$ dB, the recovery performance of the thinned $4\times5$ array in Mode 3 was not worse than Modes 1 and 2. For this experiment, Figs.~\ref{fig:sameperf} and \ref{fig:sameperf_ddm} show the plan position indicator (PPI) plot and range-azimuth-Doppler maps of all the modes, respectively. Successful detection in this case was \textit{sensu stricto}, i.e. the estimated target was at the exact range-azimuth-Doppler bin of the ground truth.\\
\noindent\textbf{Closely spaced targets}: We next considered a sparse target scene with $L = 10$ targets including two pairs of targets closely spaced in azimuth, namely, with angular spacing of $0.02$. The SNR of the injected signal was $-5$ dB. Since the angular resolution of Mode 4 is better than the other three modes, all the targets are successfully detected in Mode 4. Modes 1 and 3 produced a false alarm or missed detection as seen in the inset plots of Figs.~\ref{fig:diffperf} and \ref{fig:diffperf_ddm}. Mode 2 also shows successful recovery in the relaxed sense of our detection criterion. However, relatively better performance of Mode 2 over Modes 1 and 3 is not entirely fortuitous here. Figure ~\ref{fig:arrays} shows that both Tx and Rx array elements in Mode 2 are distributed such that its virtual array is wider than Modes 1 and 3. Thus, the effective angular resolution for Mode 2 is better than 1 and 3, but still worse than 4.\\
\noindent\textbf{Cognitive mode}: We examined a high noise scenario with an injected signal SNR = $-15$ dB. We operated only Mode 3 cognitively (i.e., with increased subband power) and kept all other modes in non-cognitive mode. We noticed that the non-cognitive Nyquist $8\times 10$ Mode 1 array exhibits false alarms while cognitive sub-Nyquist $4\times 5$ Mode 3 array is still able to detect all the targets (Figs.~\ref{fig:cogperf} and ~\ref{fig:cogperf_ddm}), thereby demonstrating robustness to low SNR. As mentioned earlier, Mode 2 and 4 continue to yield successful detections due to their wide aperture. However, contrary to Mode 2, the detection in Mode 4 is \textit{sensu stricto}.\\
\noindent\textbf{Statistical performance}: Finally, we evaluated the performance of the hardware prototype over $100$ different trials each with a distinct randomly spaced $L=10$ targets and $P=10$ pulses. We fed the target echoes to the digital receiver in all four (non-cognitive) modes as well as the cognitive Mode 3. Figure~\ref{fig:perfStats} shows the occurrence (in percentage) of specific detection results for all five configurations. We note that, as SNR worsens, the performance of sub-Nyquist, non-cognitive Mode 3 worsens in comparison with Nyquist Mode 1. However, when Mode 3 operates cognitively, its probability of correct detection increases and is actually better than Mode 1.
%-----------------------------------------------------------------------------------
%-----------------------------------------------------------------------------------
%-----------------------------------------------------------------------------------
%-----------------------------------------------------------------------------------
%-----------------------------------------------------------------------------------
%-----------------------------------------------------------------------------------
%-----------------------------------------------------------------------------------
%-----------------------------------------------------------------------------------
%-----------------------------------------------------------------------------------
%-----------------------------------------------------------------------------------
\section{Summary}
\label{sec:summary}
We presented a hardware prototype of a cognitive, sub-Nyquist MIMO radar that demonstrates real-time operation of both spatial and temporal reduction in sampling leading to reduction in antenna elements and savings in signal bandwidth. The proposed thinned 4x5 array achieves the resolution and detection performance of its filled array counterpart even though the overall reduction in bandwidth is $87.5\%$. The hardware prototype is in-house and custom-made using many off-the-shelf components. The system operates in real-time and its performance is robust to high noise. 

We demonstrated that our sub-Nyquist MIMO receiver leads to the feasibility of cognitive MIMO radar (CoSUMMeR) which transmits thinned spectrum signals. This development is significant in enabling the spectral coexistence of MIMO radar with MIMO communications. Furthermore, CoSUMMeR improves sub-Nyquist processing performance in low SNR situations by transmitting additional in-band power. We believe that such hardware implementations pave the way to identify further practical challenges in future deployment of sub-Nyquist radars.

\section*{Acknowledgements}
\label{sec:ack}
The authors are grateful to Ron Madmoni, Eran Ronen, Yana Grimovich, and Shahar Dror for hardware testing and simulations.
%\balance
\bibliographystyle{IEEEtran}
\bibliography{main}

\end{document}